\input harvmac
\input epsf
\def\lfm{\smallskip\noindent\item}
\noblackbox
\def\pl#1#2#3{{Phys. Lett. } {#1}B (#2) #3}
\def\np#1#2#3{{Nucl. Phys. } B{#1} (#2) #3}

\def\prl#1#2#3{{Phys. Rev. Lett. } {#1} (#2) #3}
\def\physrev#1#2#3{{Phys. Rev. } {#1} (#2) #3}

\def\ie{{\it i.e.}}
\def\({\left(}
\def\){\right)}
\def\[{\left[}
\def\]{\right]}
\def\ltap{\ \raise.3ex\hbox{$<$\kern-.75em\lower1ex\hbox{$\sim$}}\ }
\def\gtap{\ \raise.3ex\hbox{$>$\kern-.75em\lower1ex\hbox{$\sim$}}\ }

\def\rplus{\rho_{z_o}}
\def\rminus{\rho_{z_o+\Delta}}
\def\cp{{\it CP}}
\def\mass{{\cal M}}
\def\mc{{\cal M}_{\tilde c}}
\def\mn{{\cal M}_{\tilde n}}
\def\mst{{\cal M}_{\tilde t}}
\def\mtwoh{{\cal M}_{\tilde H}}
\def\tw{\theta_w}
\def\k{k_\perp}
\def\ie{\hbox{\it i.e.}}        
\def\eg{\hbox{\it e.g.}}        \def\cf{\hbox{\it cf.}}

\def\J{ {\cal J}}
\def\CS{ {\cal S}}
\def\H{ {\cal H}}
\def\lfm{\smallskip\noindent\item}
\def\xx2{\xi^{2}+ \xi \, {m_{i}^{2} \tau \over y T}}
\Title{ \vbox{
{\hfill UW/PT 95-07}\smallskip}}
{\vbox{\centerline{Electroweak Baryogenesis in}
\vskip6pt\centerline{ Supersymmetric Models}}}
\bigskip
\centerline{\it Patrick Huet,
 Ann E. Nelson }
\medskip
\centerline{Department of Physics }
\centerline{University of Washington, Box 351560}
\centerline{Seattle, WA 98195-1560}
\bigskip
\baselineskip 18pt
\noindent
The baryon density which may be produced during the electroweak phase
transition in supersymmetric models is computed, taking into account
the previously neglected effects of
transport, strong and weak anomalous fermion number violation,
thermal scattering, and a
new method for
computing  \cp\ violating processes during the transition. We  can
account for the observed baryon asymmetry, provided new \cp-violating
phases
are greater than $\sim 10^{-(2-4)}$, and some superpartners are
light enough to be relevant during the transition, which takes place
at a temperature of (50-100) GeV.
In one case, light superpartners are the top squarks and
the charginos and/or the neutralinos; in another case
the top squarks and both Higgs doublets are light.
Our calculation is easily extended to the case of a
general two Higgs model, where we find sufficient baryogenesis
provided that  a certain combination of parameters in the  Higgs
potential
 leads to a \cp\ violating  space dependent phase in the top quark
mass of order $ 10^{-3}$.
\Date{10/95 (revised version)}
\newsec{Introduction}
Supersymmetry is an attractive candidate for the physics of
electroweak symmetry breaking, while
electroweak baryogenesis (EWB) \ref\krs{V.A. Kuzmin,
V.A. Rubakov, M.E. Shaposhnikov, \pl{155}{1985}{36}} is an
explanation of the origin of the cosmological asymmetry between
matter and
antimatter
in terms of experimentally accessible physics. It is therefore
of interest to understand whether EWB is feasible in supersymmetric
models.
The only previous estimates\nref\dhss{ M.~Dine, P.~Huet,
R.~Singleton,~Jr. and
L.~Susskind,
   {\it Phys.\ Lett.} B 257 (1991) 351.}\nref\cn{A. Cohen and A.
Nelson,\ {\it
Phys.
\ Lett.} B297 (1992) 111. } \refs{\dhss,\cn}
of the baryon asymmetry produced in
supersymmetric models neglected many effects which are now
understood to be important, such as transport\nref\turokdiff{
M. Joyce, T. Prokopec and N. Turok, preprint
PUPT-1436 (1994), hep-ph/9401351}
\nref\ckndiffusion{A.~G.~Cohen, D.~B.~Kaplan and
A.~E.~Nelson, {\it Phys.Lett.} B336 (1994) 41, hep-ph/9406345 }
\nref\comelli{D.~ Comelli, M.~ Pietroni and A.~ Riotto, preprint
DFPD-94-TH-39, hep-ph/9406369} \refs{\turokdiff-\comelli} and thermal
scattering\nref\hs{
 P. Huet, E. Sather, hep-ph/9404302, Phys. Rev. D51 (1995)
379} \nref\us{P. Huet and A.E. Nelson,
Phys. Lett. {\bf B 355} (1995) 229, hep-ph/9504427}\nref\ghop{
M.B. Gavela, P. Hernandez, J. Orloff, O. Pene, C. Quimbay,
Mod. Phys. Lett. {\bf A}, Vol. 9, No. 9  (1994) 795;
hep-ph/9406289,
Nucl. Phys. (1994) B430 382} \refs{\hs-\ghop}.

Let us review the physics relevant for  EWB.
Anomalous  baryon violation in
the weak
interactions takes place via  unobservably slow tunnelling processes
at zero
temperature \ref\thooft{G. 't Hooft,  \prl{37}{1976}{8}
}, but at temperatures above
the critical temperature for the weak phase transition, theoretical
estimates
give a rate
$\Gamma=\kappa\alpha_w^4 T$, where $\alpha_w$  is the weak fine
structure
constant
\ref\hightemp{
P. Arnold and L. McLerran,
\physrev{D36}{1987}{581};\physrev{D37}{1988}{1020}}, and $\kappa$ is
a pure
number of order one\foot{An early estimate of $\kappa$ gave
$\kappa\gtap 0.1$\ref\kappest{J. Ambjorn, T. Askgaard, H. Porter,
M. Shaposhnikov, \pl{244}{1990}{479},\np{353}{1990}{346}} and
a recent computation claims a value
of $\kappa= 1.09\pm0.04$ \ref\ambjorn{J. Ambjorn and A. Krasnitz, preprint
NBI-HE-95-23 (1995), hep-ph/9508202}.}.
Thus electroweak baryon number violation is fast
enough in the early  universe to change the
cosmological baryon number. In thermal equilibrium, unless some
nonanomalous
approximately conserved quantum number is nonzero \ref\conserved{V.
A. Kuzmin,
V. A. Rubakov,
M. E. Shaposhnikov, Phys. Lett. 191B (1987) 171
}, anomalous processes will wash out any net
baryon number,
however a first order electroweak phase transition  can provide the
departure
from thermal
equilibrium necessary to generate a nonzero baryon number. Electroweak
baryogenesis is only
feasible if two conditions are met, which probably require new weak
scale
physics beyond the
Minimal Standard Model (MSM)\nref\cknone{A.G. Cohen, D.B. Kaplan,
A.E. Nelson,
\pl{245}{1990}{561};
 \np{349}{1991}{727}}\nref\ckn{ A.E. Nelson,
D.B. Kaplan, A.G. Cohen, \np{373}{1992}{453}}\nref\lsvt{ L.~McLerran,
M.~E.~Shaposhnikov, N.~Turok and M.~Voloshin, {\it Phys.\ Lett.} B
256 (1991)
451.}\nref\dhs{ M.~Dine, P.~Huet and R.~Singleton,~Jr, {\it
Nucl.Phys.}
B375 (1992) 625.}\nref\cknthick{ A.G. Cohen, D.B. Kaplan, A.E.
Nelson  {\it
Phys. Lett.} B263 (1991) 86.}\refs{\dhss,\cn,\cknone - \cknthick}.
  (For   relatively recent  reviews, see
\ref\rev{A. G. Cohen, D. B. Kaplan, A. E. Nelson, hep-ph/9302210,
Annu. Rev. Part. Nucl. Sci. 43
(1993) 27; P. Huet, hep-ph/9406301,
Contributed to the First
International Conference on Phenomenology of Unification
from Present to Future, Roma, ITALY - March 1994, SLAC-PUB-6492;
D.B. Kaplan, hep-ph/9503360,   Talk presented at ``Beyond the
Standard Model IV'',
Tahoe City
 12/94, University of Washington Institute for Nuclear Theory preprint
 INT95-00-86 }.)

\lfm{1.} The transition must be strongly enough first order so that
after the
transition the anomalous baryon number violation is too slow to wash
out the
baryons created
during the transition \ref\shaposh{M.E. Shaposhnikov, JETP Lett. 44
(1986)
364}. This rate is
proportional to
$\exp(-M_s/T)$, where
$M_s$, the energy of the sphaleron field configuration,  is
proportional to the
$W$ boson mass,
$M_s= (90-160) M_w$ \ref\km{F. Klinkhamer and N. Manton,
\physrev{30}{1984}{2212}}. The condition that the $W$ mass  jumps to a
large enough value during the transition to avoid post-transition
baryon number
washout requires
a  light Higgs in the MSM \nref\lighthiggs{  A.I. Bochkarev, M.E.
Shaposhnikov,
Mod. Phys. Lett. A2 (1987) 417 }
\nref\dhlll{ M.~ Dine,  P.~ Huet, R.~ G.~ Leigh, A.~
Linde and D.~ Linde, {\it Phys.\ Rev.} D 46 (1992) 550.}
\refs{\dhs,\lighthiggs,\dhlll}
(in lattice simulations, the transition
appears too weakly first order unless $m_H \ll M_w$
\ref\nonperthiggsbound{K. Farakos, K. Kajantie, K.
Rummukainen, M. Shaposhnikov,  hep-ph/9405234, Phys. Lett. B336
(1994) 494;1)
 Z. Fodor, J.
Hein, K. Jansen, A. Jaster, I. Montvay, F. Csikor, hep-lat/9405021,
Phys. Lett. B334 (1994) 405;  hep-lat/9409017, Nucl. Phys. B439
(1995) 147
}
). However with a top mass of (170-200) GeV, if the MSM is valid up to
$10^6$ GeV  we will only be
 in the MSM ground state today for a Higgs mass heavier than $\sim
M_w$
\ref\hb{
P. Arnold, S. Vokos, \physrev{44}{1991}{3620};
J.R. Espinosa, M. Quiros, hep-ph/9504241, Cern preprint CERN-TH-95-18
(1995)}.

\lfm{2.} The amount of \cp\ violation must be just right to explain
the
observed baryon to entropy ratio, $n_B/s \sim 10^{-10}$. The \cp\
violation in
the minimal standard
model is only physical in processes which involve all the
Cabbibo-Kobayashi-Maskawa (CKM) angles and in which all the like
charge quark
mass
differences play a role, which makes it seem {\it a priori} difficult
for
Kobayashi-Maskawa
\cp\ violation  to generate
sufficient baryon number during the weak transition.  An  interesting
attempt
to
find  a large enhancement of the CKM contribution to
electroweak baryogenesis  was made by Farrar and Shaposhnikov
\ref\fs{
 Glennys R. Farrar, M.E. Shaposhnikov, hep-ph/9305274,  Phys. Rev.
Lett. 70
(1993) 2833, ERRATUM-ibid.71:210,1993; hep-ph/9305275
 } but was later shown not to work due to quantum decoherence
effects \refs{\hs,\ghop}.

In constrast to the MSM case, in most extensions of the standard
model there
can be
additional sources of \cp\ violation which appear in particle mass
matrices.
During a first
order  electroweak phase transition,  bubbles of the broken phase
nucleate and expand. Inside the bubble wall, particle mass matrices
acquire
nontrivial
space-time dependence and cannot be made real and diagonal at all
points
without introducing
new \cp\ violating terms into the particle dispersion relations. In a
recent
paper
\us\ we introduced a general method for computing the effects of the
\cp\ violating mass terms on
particle distributions, which takes into
account both
the effects of
scattering from thermal particles and the  terms which lead to \cp\
violation
in
particle
propagation.   It is now established that transport
of \cp\ violating
quantum numbers into the symmetric phase,  where anomalous electroweak
baryon number
violation is relatively rapid, plays a dominant role in electroweak
baryogenesis for all
bubble wall widths \refs{\turokdiff-\comelli,\cknone,\ckn}.

The most well motivated viable theories for weak scale
baryogenesis are
two
(or more) Higgs models\ref\tz{N.~Turok, J.~Zadrozny,
Nucl.~Phys.~ B358 (1991) 471.}
 and models with weak scale supersymmetry.
In the two Higgs model the
relevant
\cp\ violation is produced by a phase in the Higgs potential, which
leads to
\cp\
violating mass matrices for fermions and Higgs bosons, and produces
especially
large
\cp\ violating effects on the Higgs and axial top number
distributions.
Experimental constraints on atomic and neutron dipole moments allow
the
relevant
phase to be as large as
$\CO(1)$
\ref\edm{see \eg \
 S.M. Barr, Int. J. Mod. Phys. A8 (1993) 209 for a review
}. Also, the two Higgs model can easily
simultaneously satisfy the constraints on Higgs particle masses and
the
requirement
of a sufficiently first order transition \nref\twohiggsphase{
 D. Land, E. D.
Carlson, Phys. Lett. B292 (1992) 107, hep-ph/9208227,  A.
Hammerschmitt,
J. Kripfganz, M.G. Schmidt, Z. Phys. C64 (1994) 105
}\nref\clinetwohiggs{ J. M. Cline, K. Kainulainen, A. P. Vischer,
preprint
MCGILL-95-16, hep-ph/9506284
}\refs{\tz,\twohiggsphase,\clinetwohiggs}. There
are
many possible supersymmetric extensions of the
 MSM,
with additional
\cp-violating phases. The minimal additional particle content (the
Minimal
Supersymmetric Standard Model, or MSSM) includes
 superpartners for all particles  and a second Higgs doublet. The
supersymmetric
terms in the Lagrangian do not introduce any additional \cp\
violation, however
supersymmetry must be broken by adding soft supersymmetry breaking
operators,
which
in general are
\cp\ violating. If the new \cp\ violating phases are of order one,
the neutron
and
atomic electric dipole moments are larger than the experimental bounds
\ref\supersymmetrycp{ J. Ellis, S. Ferrara, D.V. Nanopoulos,
Phys. Lett. 114B (1982) 231; J. Polchinski and M. B. Wise,
Phys. Lett. 125B (1983) 393; F. del Aguila, M.B. Gavela, J.A.
Grifols, A.
Mendez,  Phys. Lett. 126B (1983) 71,
ERRATUM-ibid.129B (1983) 473;  M. Dugan, B. Grinstein, L. Hall,
Nucl. Phys. B255 (1985) 413 }
unless the superpartners are unnaturally heavy
\ref\lowerbounds{  Y. Kizukuri and N. Oshimo, Phys. Rev. D45 (1992)
1806;
Phys. Rev. D46 (1992) 3025  },
hence the usual assumption is that the soft supersymmetry breaking
terms arise
from \cp\
conserving physics and have negligible phases. However it has
recently been
argued \ref\halletal{   S. Dimopoulos and L. J. Hall,   Phys.
Lett. B344 (1995) 185, hep-ph/9411273
} that in most  grand unified
supersymmetric theories, renormalization of the soft operators
between the
Planck
mass and the  scale of grand unification will induce phases of order
$10^{-2}-10^{-3}$ in the soft susy breaking operators,
providing a new source of \cp-violation into the
low energy
effective theory, which is just beyond our current experimental
reach. In this
paper
we will assume that the supersymmetry breaking terms have
\cp-violating phases
and see
whether these phases can account for sufficient baryogenesis without
violating
the
electric dipole moment bounds\ref\susyatomic{W. Fischler, S. Paban
and S. Thomas (Texas U.)  Phys. Lett. B289 (1992) 373,
hep-ph/9205233.}. In the MSSM, the requirement of a sufficiently first
order phase transition places upper limits
 on the Higgs and stop (supersymmetric partners of the top) masses
\ref\susyphase{ J.R. Espinosa, M. Quiros and F. Zwirner,
Phys. Lett. B307 (1993) 106, hep-ph/9303317;
A. Brignole, J.R. Espinosa, M. Quiros and F. Zwirner,
Phys. Lett. B324 (1994) 181, hep-ph/9312296
} which are barely consistent with experimental
constraints--there are speculations that
these bounds could be relaxed slightly by higher order and
nonperturbative effects
\ref\shap{ M. E. Shaposhnikov, Phys. Lett. B277 (1992) 324,
ERRATUM-ibid.B282:483,1992 }\foot{ Note that in supersymmetric
models,
the vacuum stability lower bounds
\hb\ on the Higgs mass do not apply.}.
The MSSM may easily be extended by adding
a gauge singlet \ref\nmssm{{\it e.g.} see
J. F. Gunion, H. E. Haber, G. L. Kane
and Sally Dawson preprint SCIPP-89/13,  (1989),  erratum
SCIPP-92-58, and refs therein} which substantially removes these
constraints. Here we will consider models
both
with and without
a singlet, but  we will only consider those sources of additional \cp\
violation
which
may be present in the MSSM, with a general set of soft supersymmetry
breaking
terms
consistent with experimental bounds. Therefore we will not
worry about
the mass
upper bounds of ref.~\susyphase, and we will assume the Higgs
potential is
\cp-conserving\foot{In some models with a singlet there can be
\cp-violation
in the
Higgs potential which can produce \cp-violating effects very similar
to those
in
two Higgs models, however in most models the Higgs potential
automatically
conserves \cp.}. We refer to the supersymmetric model  either with or
without additional
gauge singlets as the  Supersymmetric Standard Model, or SSM.

In the next
section
we discuss the dominant baryogenesis mechanisms in the SSM.
In \S 3 of this paper we write down   the set of coupled differential
equations
which describe particle interactions and transport during the weak
phase
transition,
and make reasonable approximations which allow us to find an analytic
solution
for the   baryon
asymmetry in the SSM. In \S 4 we do the same for the two Higgs model.
We
conclude with a summary of our results and their implications in \S 5.

\newsec{\cp\ violation  and
	Particle Sources in the SSM}

Following previous work
\refs{\dhss-\us,\cknone-\cknthick}  we
compute the baryon
asymmetry
using the following steps:
\lfm{I.} Compute the  \cp-violating perturbations of the plasma
locally induced
by
the passage of the wall (``particle source terms''). In ref.~\us\ we
described
all
the sources in terms of quantum mechanical \cp-violating reflection
and
transmission from layers of the
phase boundary, combined with re-thermalization of the phase-space
distributions.
 Unlike earlier
calculations, whose applicability
 was restricted to either a
``thin wall" or a  ``thick wall", referring to whether the wall
thickness is
larger or smaller than the relevant mean free paths, our approach
provides a unified and consistent treatment for all values of the wall
thickness. The proposed method links the charge generation to
microphysical CP violating processes, and hence can be widely applied.
It  generalizes the method developed in Ref.~\hs\
and so properly incorporates decoherence
effects which have been shown to have a major negative impact on the
generation
of a \cp\ violating observable in the MSM \refs{\hs,\ghop}.
\lfm{II.} We approximate the solution to the Boltzmann equations for
particle distribution functions by writing down and solving a set of
coupled
differential equations  for the local
particle densities including the source terms, transport, Debye
screening \ref\khleb{S.Yu. Khlebnikov,
\pl{300}{1993}{376}}
of induced gauge charges\foot{In practice we simplify our equations by
ignoring the
effects of screening since the impact on baryogenesis turns out to be
of order
1
\nref\debye{ A.G. Cohen, D.B. Kaplan, A.E. Nelson,
\pl{294}{1992}{57}}\nref\clinescreen{J.   M. Cline and K. Kainulainen
preprint  MCGILL-95-33, hep-ph/9506285
} \refs{\debye,\clinescreen}.}
and particle number changing reactions \ckndiffusion. The solution to
these
equations generally includes a net baryon number, which is
produced in the symmetric phase and is
transported into the bubbles of broken
phase, where it    survives until the present provided that the phase
transition is sufficiently first order.

In this section we focus on the first step in the calculation.
\cp-violating particle source terms have been shown to result
for a selected subset of species in the plasma which  mix with one
another via
a mass
matrix with complex phases which either:
\lfm{a)}   cannot be
rotated away as the result of interactions with the plasma \fs.
\lfm{b)}
cannot be rotated away at two adjacent points $x$ and $x + dx$,
by the same set of unitary transformations,
that is, $U_x^{-1}U_{x+dx} \neq 1$.

When present, the second mechanism dominates over the
first one,   as the first mechanism generically involves
additional particles whose coupling to the plasma,
yields further suppressions.
It is the second mechanism which controls baryon generation
in the SSM as
the neutralinos,  charginos, and
squarks,
have mass matrices    with \cp\ violating
entries and a non-trivial space-dependence due to the
Higgs vacuum expectations values (\cf \S 2.2).
So is the case in the two Higgs models with explicit
\cp\ violation in the Higgs potential which
yields a space dependent phase to the top quark  and Higgs
masses\foot{Another
potentially relevant species in this model is the $\tau$-lepton, and
some
have argued that its contribution dominates that of the top quark and
Higgs
\nref\taurefs{   M. Joyce, T. Prokopec and N. Turok,  Phys. Lett. B338
(1994) 269, hep-ph/9401352}\refs{\clinetwohiggs,\taurefs}. We do not
confirm the importance of the
$\tau$
lepton unless $\tan\beta$ is very large.}   (\cf \S 4).
In contrast, in the minimal
standard model, the quark mass matrix has
only an overall dependence on the Higgs vacuum expectation value
and   can be diagonalized by space-independent unitary
rotations, hence, it can generate
a \cp\ violating observable through  mechanism $a)$  rather than
through mechanism $b)$ -- \ie through charge current
interactions  which correct the dispersion relation
of the propagating quark in the plasma  \fs.
This mechanism, however, has been shown to
be quite  ineffective at generating a significant baryon asymmetry in
the MSM \refs{\hs,\ghop}.

The method introduced in Ref. \us\  can account for both mechanisms.
However, as we are concerned with extensions of the standard model
for
which the second mechanism $b)$ is dominant, we will review the
general
principles
of the method for this specific situation\foot{Our method with
mechanism $a)$, applied to the standard model,
would give results in agreement with the ones obtained in
ref. \hs, where similar
techniques have been used.}.

\subsec{The Method}

Let us consider a set of particles with mass matrix
$M(z)$
and moving, in the rest frame of the wall, with energy-momentum
$E,{\vec
k}$.
 We wish to find the \cp-violating asymmetry in their
distributions which results from their   passage
across the wall.
We define $z_o$ to be their last scattering point, where they
emerge from a thermal
ensemble with a   probability distribution
represented by a density matrix $\rplus$. These particles
propagate freely during a mean free time $\tau$,
then
rescatter and
return to the local thermal ensemble in the plane $z_o+  \tau v$,
$v$ being
the velocity perpendicular to the wall, $ k_\perp /E$.
During the time $\tau$, these
particles evolve  according to a set of
Klein-Gordon, Majorana or Dirac equations coupled through the mass
matrix
$M(z)$. (Some  effects of interaction with the plasma, which do not
destroy
quantum
coherence, can easily be included in these equations.) In the course
of this
evolution
\cp\ violation affects the distribution of these particles.
At $z_o$, the contribution of these particles to any given charge
cancel exactly the contribution of their antiparticles
since the charge  is \cp\ odd and we take the density matrix to
be \cp\ even.
However,
after evolving a time $\tau$ across the \cp\ violating space-dependent
background, this cancellation no longer takes place.
At the
subsequent scattering point $z_o+\tau v$, these charges
assume a non-zero value, as
the evolution of the particles over the distance $\tau v$ can be
\cp-violating. Specifically, the probability  for a particle
emitted at $z_0$ to be transmitted to $z_0+\tau v$ can be different
from
the transmission probability for its \cp\ conjugate.
  It is only necessary to follow the contribution of a
selected subset of charges carried by these particles in order to
characterize completely the departure from thermal equilibrium
resulting from the passage of the wall. In this subset, there are
charges which are explicitly violated by the mass matrix (axial
charge $\ldots$ ) and charges which are exactly conserved ( B-L,
$\ldots$).
In the latter case, there is no net charge density
created but, instead, a net spatial current emerges:
opposite charges
move in opposite directions.  This spatial current arises at high
order in masses and the wall velocity $v_{w}$ and, in most of the
cases we considered, yield a subleading contribution to the baryon
asymmetry. We are mostly
interested in the former charges, that is, those charges
which are violated
by the presence of the wall. In addition to a spatial current,
those charges develop a net average density;
the latter is linear in the wall
velocity and arises at low order in the mass expansion.
We will focus
our attention on those charges which develop a net  density.
We alert the reader that
there may be situations in which  conserved  charges have to be equally
considered.

To be more quantitative, we introduce $J_\pm$, the average current
resulting
from particles moving toward positive(negative) $z$ between $z_o$ and
$z_o+ \Delta $, where \eqn\deltadef{
 \Delta \equiv \tau v\ .}
The current $J_+$ receives
contributions from either
particles originating from the thermal ensemble at point $z_o$, moving
with a positive velocity and being transmitted at $z_o+ \Delta$, or
from
particles originating at $z_o+ \Delta$, moving with velocity $-v$ and
being reflected back towards $z_o+\Delta\ $ (Fig. 1a). A similar
definition
exists for $J_-$(Fig. 1b). $J_\pm$ are \cp\ violating currents
which are associated with each layer of thickness $\Delta$ moving
along with
the wall; they can be computed according to

\eqn\supersymmetryjpmgeneral{
\eqalign{
J_+ \, =& \,\bigg\{
{\rm Tr}\rplus \big[ T^\dagger{\hat Q} T \,-\, {\overline T}^\dagger
{\hat Q}
{\overline T} \big] \,  + \,
{\rm Tr}\rminus \big[ {\tilde R}^\dagger {\hat Q} {\tilde R} \,-\,
{\overline {\tilde  R}}^\dagger {\hat Q} { \overline{\tilde R}}
\big]\,
\bigg\} \pmatrix{1\cr 0 \cr 0 \cr {\tilde v }\cr} \cr
J_- \, =& \,\bigg\{
{\rm Tr}\rplus \bigg[  R^\dagger {\hat Q}  R \,-\,
{ \overline  R}^\dagger{\hat Q} { \overline  R} \bigg]\,  + \,
{\rm Tr}\rminus \bigg[  {\tilde T}^\dagger {\hat Q}  {\tilde T} \,-\,
{\overline  {\tilde T}}^\dagger {\hat Q}
{\overline {\tilde T}} \bigg]\bigg\} \, \pmatrix{1\cr 0 \cr 0 \cr -v
\cr}
\cr}
}

$R$($\overline R$) and $T$($\overline T$) are reflection and
transmission
matrices of
particles(anti-particles) produced at $z_o$ with a probability matrix
$\rplus$,
evolving
toward positive $z$ (increasing mass);
$\tilde R$ and $\tilde T$ are the corresponding quantities for
particles
produced
at $z_o + \Delta$ with probabilities contained in $\rminus$ and
evolving toward
negative
$z$;
$v$ is the magnitude of the group velocity perpendicular to
the wall at point $z_o$ while ${\tilde v}$ is the same quantity but a
distance
$\Delta$ away.
Finally, $\hat Q$ is the operator corresponding to the chosen charge
and the trace is taken over all relevant degrees of freedom and
averages over
the location
$z_o$ within a layer of thickness $\Delta$.

Formulae \supersymmetryjpmgeneral\ provide a concise method of
computing the
\cp\
violating charge currents $J_\pm$, which results from the propagation
and the
mixing of particles within the layer of thickness $\Delta$ at point
$z_o$.
After a boost to the plasma frame, these currents constitute the
fundamental
\cp\
violating building blocks that we need to construct the source terms
of
the system of rate equations introduced in \S 3, which ultimately will
convert them through diffusion and relaxation mechanisms into a net
baryon
asymmetry.

For our purpose, we construct the source terms as follows.
Consider a small volume element in the plasma. As the wall crosses
it, it
deposits
into it the current density $(J_+ + J_-)^\mu_{plasma}$ every time
interval $\tau$;
the subscript $)_{plasma}$ refers to the quantity boosted
to the plasma frame.
At an arbitrary time $t$, the current density accumulated by this
mechanism
is\foot{We
leave aside
diffusion  which is accounted for independently in the rate
equations.}

\eqn\sourcestepone{
s^\mu = \int_{t-\tau_R}^t {1 \over \tau} \, dt'\,\big(J_+(\vec x,t') +
J_-(\vec x,t') \big)^\mu_{plasma}
.}
Here, $\tau_R$ is a typical relaxation time.
  From this, we infer the net rate of change of charge $Q$ per unit
volume to be
\eqn\sourcesteptwo{\eqalign{
\gamma_Q({\vec x},t) =& \partial_\mu s^\mu \cr
 =& {1\over\tau}\,(J_+(\vec x,t) + J_-(\vec x,t))^0_{plasma}-
 {1\over\tau}\,(J_+(\vec x,t-\tau_R) + J_-(\vec x,t-\tau_R))^0_{plasma} \cr
& - \,
\int_{t-\tau_R}^t {1 \over \tau} \partial_z (J_+ + J_-)^z_{plasma}.\cr}
}

Formula \sourcestepone\ along with formulae \supersymmetryjpmgeneral\
constitute the
starting
point for our analysis of the SSM and of the two Higgs model.
In practice, we simplify eq.~\sourcesteptwo\ by making an expansion
in spatial derivatives, allowing us to neglect the third term, and we
take $\tau_R$ large, so that we can neglect the second term, whose
effect is accounted for as an independent
relaxation term in our rate equations
(cf. \S 3).

One  particular advantage  of the method above is that it
does not require any assumption about the relative magnitude of the
mean free paths  and the thickness of the wall. Hence, in contrast with
earlier methods, it unifies all electroweak baryogenesis scenarios.

\subsec{The SSM}

In the SSM, we are interested in the generation of charges which {\it
(a)} are
approximately
conserved in the unbroken phase so that, they can diffuse a long way
in front
of
the bubble wall,
where anomalous baryon violation is fast and {\it (b)} are
non-orthogonal to
baryon
number,
so that their
relaxation energetically favors a non-zero baryon charge.
Candidates of choice are Higgs number and axial top number.
The generation of these charges
results
from the mixing of the charginos, neutralinos and the mixing of
top squarks  respectively.  The chargino mass matrix is, in the basis
${\tilde W}_+,\,{\tilde W}_-, {\tilde h}_-,\,{\tilde h}_+'$

\eqn\massc{
\mc\,=\,
\pmatrix{  0&{\tilde m}_2&- v_2&0 \cr
        {\tilde m}_2&0&0&- v_1\cr
         -v_2&0&0& {e^{i\,\phi_B}}\,\mu\cr
        0&-v_1& {e^{i\,\phi_B}}\,\mu &0 \cr}
,}
with $v_1 = \,\sqrt{2} M_W(z,T)\, \cos \beta(z)$ and
	$v_2 = \,\sqrt{2} M_W(z,T) \,\sin \beta(z)$; $M_W(z,T)$ is the
temperature-dependent $W$ mass defined at each point $z$ in the wall.
The neutralinos mass matrix is, in the basis
${\tilde W}_3,\,{\tilde B}, {\tilde h}_0,\,{\tilde h}_0'$,

\eqn\massn{
\mn\,=\,
\pmatrix{  {\tilde m}_2&0&u_2 \cos \tw&- u_1 \cos \tw \cr
           0&{\tilde m}_1&- u_2 \sin \tw& u_1 \sin \tw \cr
           u_2 \cos \tw & -u_2 \sin \tw&0& -{e^{i\phi_B}}\,\mu\cr
           -u_1 \cos \tw & u_1 \sin \tw &  -{e^{i\phi_B}}\,\mu &0 \cr}
,}
with $u_1 = \, M_Z(z,T)\, \cos \beta(z)$ and
	$u_2 = \, M_Z(z,T) \,\sin \beta(z)$.
Finally, in the basis ${\tilde t}_L,\,{\tilde t}_R$,
\eqn\masss{
\mst\,=\,
\pmatrix{{\tilde m}_L^2 &  a^2 e^{i\alpha} \cr a^2 e^{-i\alpha}
&{\tilde m}_R^2 \cr}
;} here, we defined $a^2 e^{i \alpha} \,=\,
\sqrt{2}{\lambda_t \over g} \left( A\, e^{i \phi_A} v_2\, +\, \mu \,
e^{i\phi_B} v_1\right)$.

With these conventions,
the charge operator for the Higgs number
takes the form
\eqn\chargeHiggs{
{\hat Q}_{{\tilde h}} \,=\, {\rm Diag}\bigl(\, 0 \, ,\, 0 \, , \, 1
\, , \,
-1\,\bigr)
,}
while for the axial
stop number, the charge operator is defined to be
\eqn\chargestop{
{\hat Q}_{{\tilde s}} \,=\, {\rm Diag}\bigl(\, {1\over 2}\, ,\,
-{1\over 2}\,
\bigr)
.}

We can now  compute sources for those charges.
We choose to perform an expansion in powers of mass.
This expansion, introduced in \hs\ and further developed in
\us, is adequate to demonstrate the
quantum mechanical physics
required for generating a \cp\ violating observable. In particular, it
generates polynomials in ${\cal M}$ whose
imaginary part of the trace yields an expansion
in terms of \cp\ violating invariants\foot{
There are also  \cp\ violating self-energy corrections,
which are the main source of \cp\ violation in the absence of the one
considered here.}. We will discuss the validity of this approximation
later on in this section.

{\it -- Charginos and Neutralinos --}

In order to compute the source $\gamma_{\tilde h}({\vec x}, t)$
for the Higgsino number, we begin by computing the
corresponding current sources $J_\pm$ as given
in Eqs. \supersymmetryjpmgeneral.
Their determination requires the knowledge of transmission and
reflection amplitudes which are obtained by solving a
set of coupled Majorana equations with the chargino and neutralino
mass matrices given in
\massc-\massn.
We obtain up to an overall phase, at leading order in
$\mass_{({\tilde c},{\tilde n})}^{2}$,
\eqn\transHiggs{\eqalign{
T \,=\,&1 \ - \ \int_0^\Delta dz_1 \int_0^{z_1} dz_2 \mass_2 \mass^*_1
e^{i 2 \omega(z_1-z_2)} \cr
\,+\,&\int_0^\Delta dz_1 \int_0^{z_1} dz_2 \int_{z_2}^\Delta dz_3
\int_0^{z_3} dz_4
\mass_4 \ \mass^*_3 \ \mass_2 \  \mass^*_1 \  e^{i 2 \omega(z_1-z_2+z_3-z_4)}
\ +  \ \ \ldots \cr}
}
\eqn\reflHiggs{\eqalign{
{\tilde R} \,=\,& \int_0^\Delta dz_1  \mass_1
e^{-i 2 \omega z_1} \cr
\,+\,&\int_0^\Delta dz_2 \int_{z_2}^\Delta dz_3 \int_0^{z_3} dz_4
\mass_4 \ \mass^*_3 \ \mass_2  \  e^{i 2 \omega(-z_2+z_3-z_4)}
\ + \ \ \ldots \cr}
}
and similar expressions for ${\tilde T}$, ${R}$.
The quantity $\omega$ stands for the magnitude of the energy of
motion transverse to the wall, and
$\mass_i$
is short for $\mass_{({\tilde c},{\tilde n)}}(z_i)$.
The leading \cp\ violating contributions arise at order ${\cal
O}(\mass/ T)^4$.
Contributing paths are depicted in Fig.~2.

In order to use Eq. \supersymmetryjpmgeneral\ to compute $J_\pm$,
we need
the density matrices $\rplus$ and $\rminus$,
describing the distribution of particles
in phase and flavor spaces at their production point $z_o$ and
$z_o+\Delta$.
 We choose the density matrices describing equilibrium
distributions of mass eigenstates in
the unbroken phase
\eqn\densityplus{\eqalign{
\rplus^n \,=&\, {\rm Diag}\biggl(\,  n_{{\tilde m}_2}(E,{\tilde v})\,
,\,n_{{\tilde m}_1}(E,{\tilde v})\, ,\,
n_\mu(E,{\tilde v})\, ,\, n_\mu(E,{\tilde v})\,
\biggr) \cr
\rplus^c \,=&\, {\rm Diag}\biggl(\, n_{{\tilde m}_2}(E,{\tilde v})
\,,\,n_{{\tilde m}_2}(E,{\tilde v}) \, ,\,
n_\mu(E,{\tilde v}) \, ,\, n_\mu(E,{\tilde
v})\biggr) \cr}.
}
We construct $\rminus^{c,n}$ from $\rplus^{c,n}$ with the substitution
$ {\tilde v} \leftrightarrow -v$. In these expressions,
$ n_m(E,{\tilde v})$ is the Fermi-Dirac distribution, $n_f$, of a
species of
mass $m$,
boosted to the wall frame, \eqn\fermidirac{n_m=\bigl( {\rm
Exp}\bigl[\gamma_w(E-v_w
\k)\bigr]+1\bigr)^{-1}\ .}
Some motivations for our choice of density matrix are as follows.
 In a regime of large masses ${\tilde m}_2$, ${\tilde \mu} \geq T$,
it is  obviously
sensible to assume that $\rho$
is diagonal in the mass eigenstate basis;
in a regime of small masses, $\leq T$, particles are produced as interaction
eigenstates
which differ
from
mass eigenstates by a unitary rotation; ignoring this rotation
amounts to
ignoring
small corrections of order $(\mass/T)^2$.
 Furthermore, the choice of a thermal
distribution
is reasonable as the  non-equilibrium component of the distributions
is of
order $v_w$,
that is, it amounts to ignoring terms of order
$v_w^2 \, \leq 1$
\nref\mlt{ B.-H.~ Liu, L.~ McLerran, N.~ Turok, {\it  Phys.\ Rev.}
 D 46 (1992) 2668.}\refs{\dhlll,\mlt}.
Inserting Eqs. \transHiggs\ and \reflHiggs\ in
Eq. \supersymmetryjpmgeneral\ yields
\eqn\jpfour{
\eqalign{J_+ \ = \ \bigl( \,  1 \, , \ 0 \, ,\ 0 \, ,  {\tilde v}  \, \bigr)
\times \biggl\{& \cr\ +\ 4 &\int_0^\Delta dz_1 \int_0^{z_1} dz_2
\int_{z_2}^\Delta dz_3
\int_0^{z_3} dz_4 \sin 2\omega (z_4-z_3+z_2-z_1)\  \cr
\times &{\rm Im} \ {\rm Tr}
\biggl[ \rplus  \ \mass_1^* \ \mass_2 \ \mass_3^* \ \mass_4 \
{\hat Q}_{\tilde h}\biggr] \cr\ -\ 4 &\int_0^\Delta dz_1 \int_0^\Delta dz_2
\int_{z_2}^\Delta dz_3
\int_0^{z_3} dz_4 \sin 2\omega (z_4-z_3+z_2-z_1)\  \cr
\times &{\rm Im} \ {\rm Tr}
\biggl[   \ \mass_1^* \ \rminus \ \mass_2 \ \mass_3^* \ \mass_4 \
{\hat Q}_{\tilde h}\biggr] \ \  \biggr\} \cr}
}and
\eqn\jmfour{\eqalign{
J_- \ = \ \bigl( \,  1 \, , \ 0 \, ,\ 0 \, ,  {- v}  \, \bigr)
\times \biggl\{& \cr
\ -\ 4 &\int_0^\Delta dz_1 \int_{z_1}^\Delta dz_2 \int_0^{z_2} dz_3
\int_{z_3}^\Delta dz_4 \sin 2\omega (z_4-z_3+z_2-z_1)\  \cr
\times &{\rm Im} \ {\rm Tr}
\biggl[ \rminus  \ \mass_1^* \ \mass_2 \ \mass_3^* \ \mass_4 \
{\hat Q}_{\tilde h}\biggr] \cr
\ +\ 4 &\int_0^\Delta dz_1 \int_0^\Delta dz_2 \int_0^{z_2} dz_3
\int_{z_3}^\Delta dz_4 \sin 2\omega (z_4-z_3+z_2-z_1)\  \cr
\times &{\rm Im} \ {\rm Tr}
\biggl[   \ \mass_1^* \ \rplus \ \mass_2 \ \mass_3^* \ \mass_4 \
{\hat Q}_{\tilde h}\biggr] \ \  \biggr\}\, . \cr}
}
In order to proceed with  analytic expressions, we simplify further by
performing a derivative expansion: $\mass(z) \, = \, \mass (z_o) \, +
\,
(z-z_o)\, \partial_z \mass(z_o) + {\cal O}(\tau/w)^2$ and $v \, = \,
{\tilde v}
+ {\cal O}(\tau/w)^2 $.
This expansion is only justified in a region
of the parameter space for which the mean free time $\tau$ is smaller
than the
scale
of variation of the masses, i.e., the wall thickness  $w$. We will
discuss the
validity of our approximations in the last section.
This simplification
allows us to perform the trace in flavor space and in phase space
independently and yields
after summing the contributions from charginos and neutralinos,
in first order in the wall velocity $v_w$,
\eqn\jpmzandzero{
\eqalign{
(J_+ + J_-)^z  =& \ 0 \cr
(J_+ + J_-)^0   =& \ \gamma_w v_w \, \times \,  \sum_i^{\rm eigen.} \
\J^i \ \times \ \int {d^3 {\vec k}\over (2\pi)^3}
 \ {f(\omega_{i} \Delta)\over \omega_{i}^5 } \ (2 v_i)\ {e^{E_i/T}
\over (1+ e^{E_i/T})^2} {E_i\over T} \cr
}
}
with $f(\xi)$ defined as
\eqn\fdef{ f(\xi)\,=\, (\xi\, \cos \xi \,-\,
\sin \xi )^2
}
and the summation is over  the mass eigenstates with eigenvalues $m_i$
in the unbroken phase, with
$\J^i$,  a corresponding \cp-violating invariant. For the charginos,
there are two mass eigenstates with masses $\mu$ and $ {\tilde
m}_2$,  whose
\cp-violating invariants are
\eqn\jcpcharginos{\eqalign{
\J_{ch}^\mu \, =& \J_{ch}^{{\tilde m}_2} \, = \,
  - \ \sin \phi_B \ \mu \  {\tilde m}_2 \ ( v_2\partial_z v_1 \,-\,
v_1\partial_z v_2 )/T^5  \cr
=&\, 2\ \mu \  {\tilde m}_2 \ M_W^2(T,z) \ \sin
\phi_B \
\partial_z \beta  \ / T^5   \, . \cr}
}
While, for the neutralinos, there are four eigenstates with masses
$\mu$, $\mu$, $ {\tilde m}_1$ and $ {\tilde m}_2$ and with,
correspondingly,
\eqn\jcpneutralinos{\eqalign{
\J_{n}^\mu \, =& (\J_{n}^{ {\tilde m}_1} + \J_{n}^{ {\tilde m}_2})/2
\cr
 \J_{n}^{ {\tilde m}_1} \,=&  \, -\ \sin \phi_B  \
\mu \ {\tilde m}_1 \sin^2 \tw \ ( u_2\partial_z u_1 \,-\,
u_1\partial_z u_2
)/T^5
\cr
=&\,  \mu \  {\tilde m}_1 \ M_Z^2(T,z) \sin^2 \tw  \
\sin \phi_B \
\partial_z \beta \ / T^5\cr
 \J_{n}^{ {\tilde m}_2} \,=&  \, - \ \sin \phi_B  \
\mu \ {\tilde m}_2 \cos^2 \tw \ ( u_2\partial_z u_1 \,-\,
u_1\partial_z u_2
)/T^5
\cr
=&\,  \mu \  {\tilde m}_2 \ M_Z^2(T,z) \cos^2 \tw  \
\sin \phi_B \
\partial_z \beta \ / T^5  \, . \cr}
}

We can now construct the local source density
$\gamma_{\tilde h}({\vec x}, t)$ for the Higgs number by inserting Eq.
\jpmzandzero\
in formula \sourcesteptwo, we obtain
\eqn\Higgsinos{\eqalign{
\gamma_{\tilde h}({\vec x}, t) \, = \, \gamma_w v_w \, {T^4 \over 4
\pi^2}  \,\bigl(\,
\J_{ch}^{  \mu}   \,
{\cal I}_{\tilde h}^{\mu}  +&  \J_{ch}^{ {\tilde m}_2}  \,
{\cal I}_{\tilde h}^{{\tilde m}_2}+ 2\, \J_{n}^{ \mu}   \,
{\cal I}_{\tilde h}^{\mu}+         \J_{n}^{ {\tilde m}_1}  \,
{\cal I}_{\tilde h}^{{\tilde m}_1}+ \J_{n}^{ {\tilde m}_2} \,
{\cal I}_{\tilde h}^{{\tilde m}_2} \bigr) \cr
+& {\cal O}\bigl(  v_w^2, (\tau/w)^2\bigr) \cr}
}

The factor ${\cal I}_{\tilde h}^{m}$ contains information on
the
phase space as well as on effects due to plasma interactions. Its
analytic form is
\eqn\formHiggs{
{\cal I}_{\tilde h}^m \, = \, \sqrt{ \tau T} \ \int_{m \over T}^\infty
dy\ y^2 \ {e^y \over (1+e^y)^2} \
\int_0^{\tau T (y^2-({m \over T})^2)/y} d \xi \
{f\left(\sqrt{\xx2}\right) \over \left( \xi y +
{m^2 \tau \over T}\right)^{5/2}} }
it is a simple exercise to show that
the   factor ${\cal I}_{\tilde h}^m$ vanishes rapidly with $\tau$, as
\eqn\hsmalltau{
{\cal I}_{\tilde h}^m \, \propto \ \tau^5 \qquad \qquad (\, \tau \to
0 \, ).
}
This steep dependence on $\tau$, simply reflects the high suppression
resulting from decoherence due to incoherent scatterings in the
plasma whose frequency
increases as $1/\tau$. This suppression has the same origin as the
one forbidding electroweak baryogenesis with \cp\ violation
originating
from the mixing of light quarks
in the MSM.
For larger coherence time, ${\cal I}_{\tilde h}^m$ scales
approximately
as
\eqn\hlargetau{
{\cal I}_{\tilde h}^m \, \propto \,  {1 \over \tau T}
 \qquad \qquad (\, \tau \to
\infty \, ).
}

This fall-off with increasing $\tau$ describes the semi-classical limit
in which particles propagate a distance long compared to their
Compton wavelength in which case, fast oscillations of their
wave-function wash away the interference required to generate a $\cp$
asymmetry.
This fall-off takes place in the thick wall
regime, $\tau \leq w$, the situation for which our derivative
expansion
applies; it persists until $\tau \simeq w$,
which defines the thin wall regime.
The two behaviors \hsmalltau\ and \hlargetau\ are easily identified in Fig.~3.

For the case of interest, the damping rate is
essentially dominated by weak interaction processes:
$\gamma_{\tilde h}  \,
\sim \,
\alpha_w  T$.
A  crude estimate for the coherence time $\tau_ {\tilde h}$,
$\,\sim \gamma_{\tilde h}^{-1}$,
 yields the range
$ 20/T \, \leq \,\tau_{\tilde h} \,\leq 30/T$, which lies comfortably
in the asymptotic domain
described by \hlargetau.

In summary, an analytic form which
fits well the source term for the Higgsino number
in the domain of interest is Eq. \Higgsinos\ with
\eqn\hanalytic{
{\cal I}_h^m \, \approx\,  {22 \over 5 } \ {1\over T\tau_{\tilde h}}
\ {e^{m/T} \over (1+e^{m/T})^2}\left({T \over m}\right)^{5/2} \qquad
20/T \,\leq \,\tau_{\tilde h} \,\leq 30/T \ .}
This form is  only a fit which is valid in the range  $0.5 < m  /T \leq
1$.

Let us assess the domain of validity of the expression above.

\lfm{1.} Results \formHiggs-\hlargetau\ are not to be trusted in
the limits $\tau > w$ where
our  derivative  expansion does not apply. Instead, in this limit,
one expects to observe the
dependence on $\tau$ to weaken as $\tau \to w$, and to vanish as
$\tau \gg w$;
that is,
one expects to observe the  factor $ {\cal I}_h^m $ to saturate at a
value
$\tau \sim w$,
to become a function of the mass only.

\lfm{2.} The coherence time
$\tau_{\tilde s}$ given above is an estimate based on our
knowledge
on the damping rate
which, so far, has only been studied in the low momentum
limit\ref\damping{
 E.~Braaten and R.~D.~Pisarski, {\it Phys. Rev.} {\bf D}
 46 (1992) 1829; see also F.~Flechsig, A.~K.~Rebhan, H.~Schulz,
hep-ph/9502324,  DESY 95-022, ITP-UH-06/95 and references therein.};
this value can only be a {\it crude} estimate.

\lfm{3.}
Finally, we turn to the important question of the validity of the
mass expansion we used to derive formulas \Higgsinos-\hlargetau.
A careful study of this expansion shows that its expansion parameter
 is either $m \Delta $ in the limit $\omega\Delta  \leq 1$
or
$m /\omega$ in the limit $ \omega \Delta\geq 1$. In both
limits,
the expansion parameter is less than one because of the relation
$m < \omega = \sqrt{\k^{2} + m^{2}}$.  Let us give the physical
interpretation of those statements\foot{For a more detailed discussion see
Ref. \hs.}. In a typical scattering off a
diffracting medium characterized by
a step potential of height $V$, reflection and transmission
amplitudes result from the constructive and destructive interference
of various diffracted waves generated everywhere
inside the bulk of the medium.
 Transmission and reflection amplitudes will be
comparable\foot{In order to
obtain  significant \cp\ violating
asymmetries, both  reflection and transmission amplitudes are to be
significantly different from zero, otherwise, as either $|R|$ or
$|T|$ goes to zero, the other one goes to one from unitarity, and
both $|R|^2-|{\overline R}|^2$ and $|T|^2
-|{\overline T}|^2$ vanish correspondingly.}
if the incoming
wave penetrates {\it coherently} the diffracting medium over at least
a distance of order $1/V$, and has few oscillations over
that distance.
 Suppression of the
reflection amplitude
arises if the coherence length $\Delta$ of the incoming
wave is smaller
than $1/V$ or if its energy $\omega$ is larger than $V$.
In the case where $\Delta \ll 1/V$,
only a layer $\Delta$ of the medium effectively contributes
to the coherent reconstruction of the reflection and
transmission amplitudes, this is the phenomenon of decoherence, which
suppresses the reflection amplitude with powers of  $\Delta \  V$. In the
case where $V \ll \omega$,
fast oscillations of the propagating wave inside the medium,
tend to attenuate the reconstruction of the reflected amplitude with
powers of $V/\omega$. In the present case, the ``diffracting medium''
is the wall and its height $V$ is the mass $m$ of the scattering particle,
hence, suppression factors are controlled by $\Delta \  m$ or $m/ \omega$,
 whichever is smaller. An alternative
method of computation of the currents
$J_\pm$ consists of
computing $R$ and $T$ by  solving
Majorana equations including the
imaginary
part of the thermal self-energy. This method automatically accounts
for both effects occuring here \refs{\hs}. In particular, use of the mass
expansion in this context suggests an  expansion parameter
$\sim m\Delta/\sqrt{1+\omega^{2}\Delta^{2 }}$,  which corroborates
 the analysis above.

{\it -- Squarks --}

We now turn to the calculation of the source for the  axial
stop number.
The stop mass matrix is given in \masss\ and the top axial charge
operator
${\hat Q}_{\tilde s}$ is given in \chargestop.
As for the Higgsino number, we proceed in computing the current source
$J_\pm$ in
the wall
frame using Eqs. \supersymmetryjpmgeneral, which we then input
into formula \sourcesteptwo\
to construct the source $\gamma_{\tilde s}({\vec x}, t)$.

This time, the amplitudes are computed in solving a set of coupled
Klein-Gordon
equations.
We obtain up to an overall phase, at leading order in $\mst^2$,

\eqn\transstop{
T \,=\,  \bigg[1 \ + \dots\ - \
\int_0^\Delta dz_1 \int_0^{z_1} dz_2
{\mass^2_2 \over 2  \omega}\, {\mass^2_1 \over 2  \omega}\ e^{i 2
\omega(z_1-z_2)} \bigg]
\ + \ \ \ldots
}and
\eqn\reflstop{
{\tilde R} \,=\, \dots \, + \,
\int_0^\Delta dz_1
{\mass^2_1 \over 2  \omega}\ e^{-i 2 \omega z_1}
\ + \ \ \ldots
}
We only displayed the contributions whose interference contribute to a
\cp\ asymmetry in $J_+$. These specific paths are depicted
in Fig.~4.
As we did earlier, we assume the squark density matrices $\rplus$ and
$\rminus$
to describe thermal distributions in the unbroken phase
\eqn
\densitysquark{\eqalign{
\rplus^s \,&=\, {\rm Diag}\biggl(\,  n_{{\tilde m}_L}^b(E,+{\tilde
v})\, ,\,
                               n^b_{{\tilde m}_R}(E,+{\tilde  v})\,
\biggr) \cr
\rminus^s \,&=\, {\rm Diag}\biggl(\,  n^b_{{\tilde m}_L}(E,- v)\, ,\,
                               n^b_{{\tilde m}_R}(E,- v)\, \biggr)
\cr}
.}

$ n^b_m(E,{\tilde v})$ now, refers to the Bose-Einstein distribution,
$n_b$,
boosted to the wall frame, \eqn\bosee{
n^b_m=\bigl( {\rm Exp}\bigl[\gamma_w(E-v_w
\k)\bigr]-1\bigr)^{-1}.}
The soft supersymmetry  breaking masses are kept in \densitysquark\
as, for large values, they yield an exponential suppression of the
baryon
asymmetry produced.

 From \transstop, we obtain for the current sources
\eqn\jptwo{\eqalign{
J_+ \,= \,\bigl( \, 1 \, , \ 0 \, ,\ 0 \, ,   {\tilde v}
\ \bigr)  &\times  \bigg\{ \cr
	+&\int_0^\Delta dz_1 \int_0^{z_1} dz_2 \
{\sin 2\omega (z_1-z_2)\over \omega^2}
\times {\rm Im} \ {\rm Tr}\biggl[ \rplus \  {\hat Q}_{\tilde s} \
\mass^2_2 \
\mass^2_1 \,
\biggr] \cr
	+&\int_0^\Delta dz_1 \int_0^\Delta dz_2 \
{\sin 2\omega (z_1-z_2)\over 2 \omega^2}
\times {\rm Im} \ {\rm Tr}\biggl[ \rminus \ \mass^2_2
\  {\hat Q}_{\tilde s} \  \mass^2_1 \,
\biggr]  \ \ \bigg\}
\cr}} and
\eqn\jmtwo{J_- \, = \, J_+( \, \rplus\, \leftrightarrow\, \rminus , \,
{\tilde v} \, \leftrightarrow\, -v \,).
}

Performing an expansion in the wall velocity $v_{w}$, we find, in first
order in $v_w$,

\eqn\jpmzandzerosquarks{
\eqalign{
(J_+ + J_-)^z  =& \, 0 \cr
(J_+ + J_-)^0   =& \, \gamma_w v_w \  \times  \J_{\tilde s}
 \times
\sum_{i={\tilde m}_L,{\tilde m}_R} \
 \int {d^3 {\vec k}\over (2\pi)^3}
 \ {g(\omega \Delta)\over 4 \omega^5 } \ (2 v_i)\ {e^{E_i/T}
\over (1- e^{E_i/T})^2} {E_i\over T} \cr}
}
where $g(\xi)$  is defined as
\eqn\xgdef{ g(\xi)\,=\, 1 \, - \, \cos 2 \xi \, -\, \xi\, \sin 2 \xi
}
and
$\J_ {\tilde s}$ is a new \cp\ violating invariant given
by

\eqn\jcps{\eqalign{
\J_{\tilde s} \, =& \,
- \partial_z \alpha \ a^4/T^5 \cr
=& \, 4 {\lambda_t^2 \over g^2} \, A\, \mu\, \sin(\phi_B-\phi_A)\,
M_W^2(T,z)\,
\partial_z \beta \ .}
}

After a few simple manipulations, we derive the following expression
for
the stop axial source
$\gamma_{\tilde s} ({\vec x}, t)$,
\eqn\squarks{\eqalign{
\gamma_{\tilde s }({\vec x}, t)\, = \, \gamma_w v_w \ &N_c \ {T^4
\over 4 \pi^2}
 \,
 \times
\J_{\tilde s}   \times
\bigl({\cal I}_{\tilde s}^{{\tilde m}_L} \  + \ {\cal I}_ {\tilde
s}^{{\tilde
m}_R} \bigr) \cr \, &+ \ {\cal O}\bigl(  v_w^2, (\tau/w)^2\bigr) \cr}
}
where $N_c$ is the number of colors, $=3$.
The   function  ${\cal I}_{\tilde s}^m$ is given by
\eqn\formsquarks{
{\cal I}_{\tilde s}^m \, = \, {1\over 4} \  \sqrt{ \tau T} \ \int_{m
\over
T}^\infty
dy \ y^2 \ {e^y \over (1-e^y)^2} \
\int_0^{\tau T (y^2-({m \over T})^2)/y} d \xi \
{g\left(\sqrt{\xx2}\right) \over
\left(\xi y + {m^2 \tau \over T}\right)^{5/2}}
.}
it is simple to show that
the  factor ${\cal I}_{\tilde s}^m$ vanishes rapidly with $\tau$
\eqn\ssmalltau{
{\cal I}_{\tilde s}^m \, \propto \ \tau^3 \qquad \qquad (\, \tau \to
0 \, ) \ ,
}
as, in this limit, incoherent plasma scatterings become overwhelming.
This behavior, already noted in the Higgsino case, is a universal
property which can be traced to the quantum nature
of \cp\ violation, conflicting with the classical nature of the
plasma physics.
For larger coherence time, ${\cal I}_{\tilde s}^m$ behaves
approximately as\foot{This behavior is cut off at a value $\tau \sim
L$; at this value, ${\cal I}_{\tilde s}^m$ is expected to saturate
to its ``thin wall'' value.}
\eqn\slargetau{
{\cal I}_{\tilde s}^m \, \propto \,  {1 \over \tau T}
\qquad \qquad (\, \tau \to
\infty \, ).
}
Both behaviors \ssmalltau\ and \slargetau\ are evident on Fig.~5.
Unlike the case of the Higgsinos, the squark plasma physics is
dominated by
strong
interactions, and so is the damping rate, hence  we estimate
$\tau_{\tilde s},\,\propto \ (2 \alpha_s T)^{-1}$, to be about
$5/T$. From
this estimate, we infer that the regime of relevance is neither
the decoherence regime nor the semi-classical regime but rather
an intermediate regime corresponding to the peak shown on Fig. 5. This is a
situation  already encountered in Ref. \us, in the case of the
top quark in the two Higgs model (cf. \S 4).

In summary, and with the above value for $\tau_{\tilde s}$,
an analytic expression which provides
a reliable fit to
the source $\gamma_ {\tilde s} ({\vec x}, t)$ for the axial stop
number,
in the  range $0.5 \leq m/T $ , is\foot{We emphasize that this is {\it only} a
fit.}
\eqn\squarksanaly{
\gamma_{\tilde s}({\vec x}, t) \, \simeq \, \gamma_w v_w \ { N_c T^4
\over 200
\pi^2}  \
\times \ \J_{\tilde s}  \ \times \ \sum_{m={\tilde m}_L,{\tilde m}_R}
{T \over m}   \ {e^{m/T}\over(1-e^{m/T})^2}
 \qquad
{\rm with}\qquad \tau_{\tilde s} \
\simeq \ 5/T \ .
}
 This expression has been derived under the same
assumptions as the ones made to derive the corresponding
analytic form \hanalytic\ for the source for the
Higgsino number, $\gamma_{\tilde h} ({\vec x}, t)$.
These assumptions have been
evaluated in the discussion following Eq. \hanalytic.

\newsec{Diffusion equations in the SSM}

 Only those particle species which participate in particle number
changing
transitions which
are fast compared with  the relevant timescales, but which carry some
charge
which
is approximately conserved in the symmetric phase, can have
significant nonzero
densities
in the symmetric phase during the transition.  If the system is near
thermal
equilibrium and the
particles interact weakly, the particle densities
$n_i$ satisfy
\eqn\thermaldensities{n_i = k_i\mu_i T^2/6\ ,}
where $\mu_i$ is a local chemical potential for particle species
$i$,  and
$k_i$ is a
statistical factor defined by eq.~\thermaldensities. For light, weakly
interacting particles
$k_i\approx 2$(boson degrees of freedom)$+1$(fermion degrees of
freedom), while
for particles
much heavier than $T$ it is exponentially small. If we consider a
reaction
which changes the
 particle number of particle species $i$ by $\Delta_i$, near thermal
equilibrium the
difference beween the rates for the reaction and its inverse will
satisfy
\eqn\react{
\Gamma_{\rm net}={\sum_i\Delta_i\mu_i\over T} \Gamma_{\rm
fluct}={\sum_i
n_i\Delta_i\over k_i}
6\Gamma_{\rm fluct}/T^3\ ,} where $\Gamma_{\rm fluct}$ is the total
rate for
the reaction and
its inverse per unit volume. For convenience   we will henceforth
define
particle number
changing rates   to be
$(6/T^3)\Gamma_{\rm fluct} $.

 We can now write down a set of coupled
differential equations which include the effects of diffusion,
particle number
changing
reactions, and \cp\ violating source terms, and solve them  to find
the various
particle
densities in the SSM.   Anticipating a small departure from
equilibrium, we incorporate particle number changing
reactions and sources as two distinct terms. Diffusion is described  by a
standard diffusion term without a provision to account for the potentially
fast relative motion of the sources in respect to the plasma.  It is a good
description in the regime of a wall velocity $v_{w}$ small compared to the
speed of sound in the plasma $c_{s} = 1/\sqrt{3}$. This condition,
which is likely to be fulfilled in the minimal standard model
\refs{\dhlll,\mlt},
may or may not be fulfilled
in more general theories such as the ones considered here.
To find out would require a complete calculation of the phase transition,
which is beyond the scope of the present work.
Further simplifications of these equations take
place when we neglect all
couplings except for  gauge interactions, and the top quark Yukawa
coupling. We
include the
effects of strong sphalerons\nref\moh{
 Rabindra N. Mohapatra, Xin-min Zhang, Phys .Rev. D45 (1992)
2699}\nref\giudice{G. Giudice and
M. Shaposhnikov,
\pl{326}{1994}{118} } \refs{\moh,\giudice}, but neglect the weak
sphalerons
until near the end
of the calculation. The neglect of the weak sphalerons allows us to
forget
about
leptons in our differential equations, and will turn out to be a good
approximation
when computing Higgs and quark densities. We also neglect the effects
of
hypercharge
gauge forces and screening, which can be shown   to   affect the
baryon
number produced by  a factor  of at most order one  \clinescreen.
The  particle
densities we need include
$q\equiv (t_L+b_L)$, the right handed top quark
$t\equiv t_R$,   the Higgs particles $h\equiv (h^- + h^0 +
\bar h'^+ +
\bar h'^0)$, and their superpartners $\tilde q, \tilde t, \tilde h$.
The
individual particle numbers of these species can change  through the
top quark
Yukawa
interaction,  the top quark mass, the Higgs self interactions, and
anomalous weak interactions, and the supergauge interactions. We will
find that
baryogenesis
in the minimal model is only feasible if some of the superpartners of
the gauge
and Higgs
bosons are light, so that we may take the supergauge interactions to
be in
thermal
equilibrium ($q/k_q=\tilde q/k_{\tilde q},t/k_t=\tilde t/k_{\tilde t},
h/k_h=\tilde h/k_{\tilde h}$), and describe the system by densities
$Q=q+\tilde q,\ T =t+\tilde t$ and
$H=h+\tilde h$.
 As shown in \S~2,   \cp\ violating interactions with the phase
boundary
produce source terms $\gamma_{\tilde h}$ for the Higgsinos and
$\gamma_{\tilde
s}$ for the
 $\tilde q-\tilde t$
densities, which tend to pull the system away from equilibrium. When
we include
strong
sphalerons (with a rate
$\Gamma_{ss}$), we will
  generate   a right handed bottom
quark   density,
$B\equiv b_R+\tilde b_R$,  as well as first and second family quarks
$Q_{(1,2)L}, \  ,U_R \  ,C_R \  ,S_R \  ,D_R$.
 However since  strong sphalerons are the only processes which
generate
significant numbers of
first and second family quarks, and  all quarks  have nearly the same
diffusion constant,  we can constrain these densities   algebraically
in terms of $B$
to satisfy\eqn\algebra{Q_{1L}=Q_{2L}=-2U_R=-2D_R=-2S_R=-2C_R =
-2B=2(Q+T)\
.}
   For simplicity we will also assume all squark partners of the light
quarks are degenerate and take
\eqn\degsquark{k_{Q_{1L}}=k_{Q_{2L}}=2k_{S_R}=2k_{D_R}=
2k_{U_R}=2k_{C_R}=2k_B\
.}
 We  include
scattering processes involving the top quark Yukawa coupling, with
rate
$\Gamma_{y}$, and in the phase boundary and broken phase we have Higgs
violating
processes  at a rate $\Gamma_h$ and axial top number violation at a
rate
$\Gamma_m$.
Following ref.~\ckndiffusion, particle transport is treated by
including a
diffusion term. We
take all the quarks and squarks to have the same diffusion constant
$D_q$ and the Higgs and Higgsinos to have diffusion constant $D_h$.

The rates of change of the various densities are now described by the
coupled
equations:
\eqn\qth{\eqalign{
\dot Q &= D_q \nabla^2 Q - \Gamma_y\[{Q/ k_Q} -{H/ k_H} -
{T/ k_T}\]  -\Gamma_m\[{Q/ k_Q} -{T/k_T} \] \cr
&\qquad
 - 6 \Gamma_{ss}\[2Q/ k_Q - T/ k_T+9 (Q+T)/k_B\]+\gamma_{\tilde s}\cr
 \dot T &= D_q \nabla^2 T  - \Gamma_y\[-{Q/ k_Q} +{H/ k_H}
+ {T/ k_T}\]  \cr
&\qquad -\Gamma_m\[-{Q/ k_Q} +{T/ k_T}\] +
3 \Gamma_{ss}\[2Q/ k_Q - T/ k_T+9 (Q+T)/k_B\] -\gamma_{\tilde s}\cr
\dot H &= D_h \nabla^2 H - \Gamma_y\[-{Q/ k_Q} +{T/ k_T}
+ H/k_H\] -\Gamma_h{H/ k_H}+\gamma_{\tilde h}
 \ .\cr }}
Several simplifications of equations \qth\ can be made.
First we ignore the curvature of the bubble wall,
and so $\Gamma_m$, $\Gamma_h$,
and $\gamma_{\tilde s,\tilde h}$
are only functions of $\bar z\equiv\vert\vec r+\vec v_wt\vert$,
where  $\vec v_w$ is the bubble wall velocity.
We will  assume that the density perturbations of interest are
only functions
of $\bar z $, the coordinate normal to the wall
surface.

With these assumptions we arrive at the  equations for
$Q(\bar z),\ T(\bar z)$,  and $H(\bar z)$ in
the rest frame of the bubble wall:

\eqn\masteq{\eqalign{
0 &= -v_w Q' + D_q Q'' - \Gamma_y\[Q/k_Q -H/k_H - T/k_T\]
-\Gamma_m\[Q/k_Q
-T/k_T \] \cr
&\qquad     - 6 \Gamma_{ss}\[2Q/ k_Q - T/ k_T+9
(Q+T)/k_B\]+\gamma_{\tilde
s}\cr
0&=-v_wT' + D_qT'' + - \Gamma_y\[-{Q/ k_Q} +{H/ k_H}
+ {T/ k_T}\]  \cr
&\qquad -\Gamma_m\[-{Q/ k_Q} +{T/ k_T}\] +
3 \Gamma_{ss}\[2Q/ k_Q - T/ k_T+9 (Q+T)/k_B\] -\gamma_{\tilde s}\cr
0 &=-v_wH'  + D_h H'' - \Gamma_y\[-{Q/ k_Q} +{T/ k_T}
+ H/k_H\] -\Gamma_h{H/ k_H}+\gamma_{\tilde h}\ .\cr}}

We now assume that the rates $\Gamma_y$ and $\Gamma_{ss}$ are fast,
and so
$Q/k_Q -H/k_H - T/k_T
=\CO(1/\Gamma_y), 2Q/ k_Q - T/ k_T+9 (Q+T)/k_B=\CO(1/\Gamma_{ss})$.
We will
check later whether
this assumption is self consistent.   We then take the linear
combination of
eqs.~\masteq\
which is independent of
$\Gamma_{ss},\Gamma_y$, and substitute
\eqn\qtsol{\eqalign{Q= &H\left({k_Q(9k_T-k_B)\over k_H(k_B+9
k_Q+9k_T)}
\right)+\CO(1/\Gamma_{ss}, 1/\Gamma_y)\cr
T=& -H \left({k_T(2k_B+9k_Q)\over k_H(k_B+9 k_Q+9
k_T)}
\right)+\CO(1/\Gamma_{ss}, 1/\Gamma_y)\ .}} We then find that the
Higgs density
satisfies\eqn\Higgsdens{ 0=-v_w H' +\bar D H''-\bar\Gamma
H+\bar\gamma+\CO(1/\Gamma_{ss},
1/\Gamma_y)} where $\bar D$ is an effective diffusion constant,
$\bar\Gamma$
is
an
effective decay constant and $\bar\gamma$ is an effective source
term, given by
\eqn\effdconst{\eqalign{\bar D=
&{D_q(9k_Qk_T-2k_Qk_B-2k_Bk_T) +D_Hk_H(9k_Q+9k_T+k_B)
\over 9k_Qk_T-2k_Qk_B-2k_Bk_T+k_H(9k_Q+9k_T+k_B)}\cr
\bar\gamma=&(\gamma_{\tilde s}+\gamma_{\tilde
h})\left({k_H(9k_Q+9k_T+k_B)\over
9k_Qk_T-2k_Qk_B-2k_Bk_T+k_H(9k_Q+9k_T+k_B)}\right)\cr
\bar\Gamma=&(\Gamma_m+\Gamma_h)\left({9k_Q+9k_T+k_B\over
9k_Qk_T-2k_Qk_B-2k_Bk_T+k_H(9k_Q+9k_T+k_B)}\right)\ .}}

In these equations, $\bar\gamma$ is the sum of the rate
of generation of axial quark number and Higgs number inside the wall
as
given in \squarks\ and \Higgsinos\ while, $\bar\Gamma$ is the total
rate
of relaxation for those charges. We estimate the latter to be
\eqn\hyper{
(\Gamma_m+\Gamma_h) \,\approx\, {4 M^2_W(T,z) \over 21 g^2 T}\
\lambda_t^2 \ \sin^2 \beta \ + \
 { M^2_W(T,z) \over 35 g^2 T}
\, . }

Equation \Higgsdens\ is easily solved numerically for arbitrary shape
of the
source
$\bar \gamma$ and
decay term $\bar\Gamma$, however in order to qualitatively understand
how the
baryon number
produced depends on the various parameters we will approximate the
source as a
step function of
width
$w$
 \eqn\approxsource{\eqalign{\bar\gamma=& \, \tilde\gamma,\quad   w >
\bar z
>0\cr
                             \bar\gamma=&\,0,\quad \bar z >  w,\ \bar
z < 0\
,}}
while for the decay terms we take
\eqn\approxdecay{\eqalign{\bar\Gamma=&\,\tilde\Gamma,\quad \bar z>0\cr
                          \bar\Gamma=&\,0,\quad \bar z < 0\ .\cr}}
The effective diffusion constant is also  spatially varying since the
statistical factors $k_i$ depend on spatially varying particle masses
and since
the weak
interaction cross sections depend on the Higgs vevs, however we will
make the
reasonable
approximation that
$\bar D$ is  constant. An analytic solution to eq.~\Higgsdens, which
satisfies
the boundary
conditions
$H(\pm\infty)=0$ is now readily found; for $\bar z<0$ (the symmetric
phase)
this is
\eqn\sol{H={\cal A}\,e^{ \bar z v_w/\bar D }} with
\eqn\ampl{{\cal A}=\,{4\tilde\gamma {\bar D} \left(
1-e^{-\left[(v_w+\sqrt{4 \bar D \tilde\Gamma+v_w^2})( \bar
D)\right]}\right)\over
\left(v_w+\sqrt{4
\bar D
\tilde\Gamma+v_w^2}\right)^2 } \ .}
We will see that $\bar D\tilde\Gamma\gg v_w^2$ and so a good
approximation to
eq.~\ampl\ is
\eqn\approxampl{\eqalign{
{\cal A}\approx&\left({\tilde\gamma \over
\tilde\Gamma}\right)\left(1-e^{-2 w
\sqrt{\tilde\Gamma\over\bar D}}\right) \cr
	\approx& \ k_H \ \left({\gamma_{\tilde s} +\gamma_{\tilde h} \over
	\Gamma_m + \Gamma_y}\right)\left(1-e^{-2 w
\sqrt{\tilde\Gamma\over\bar D}}\right)\ . \cr
}}
  From the form of \sol, we see that the \cp\ violating densities are
non zero
for
a time $t =
\bar D/v_w^2$, and so the assumptions about which rates are fast
which were
used to derive
eq.~\Higgsdens\ are valid provided
$\bar D \Gamma_{ss}/v_w^2, \bar D \Gamma_y/v_w^2 \gg1$, $\bar D
\Gamma_{ws}/v_w^2\ll 1$,
and the scattering processes due to Yukawa couplings other than top
are slow.

To estimate $\bar D$ we take the Higgs diffusion constant
$D_h$ to be comparable to the diffusion constant for left handed
leptons, which
was estimated
in the MSM in Ref.~\ref\jptthin{M. Joyce, T. Prokopec, N. Turok,
PUPT-1495
(1994) hep-ph/9410281} to be
$110/T$ and
take $D_q$ from \jptthin\ to be $6/T$. (These numbers will decrease
slightly
due to the
supersymmetric particle content of the plasma--we ignore this  effect
as being
small compared
with other uncertainties in our calculation.) For the
$k_i$'s we assume that all the supersymmetric particles are heavy
compared with
$T$ except for
the neutralinos and charginos and so \eqn\counting{k_Q\approx6,\quad
k_T\approx
3,\quad
k_B\approx 3,\quad k_H\approx 12\ .} We then find the effective
diffusion
constant defined in
eq.~\effdconst\ is
large, \eqn\diffbar{\bar D\approx 100/T\ .} The large effective
diffusion
constant indicates that most of the transport of
\cp\ violating quantum numbers is done by weakly interacting
particles, \ie the
Higgs and Higgsinos, and since Yukawa interactions readily convert
Higgs number
into axial top number, transport of axial top number is surprisingly
efficient.

For the   scattering rate due to the top quark Yukawa coupling we
estimate
\eqn\gamy{\Gamma_y\approx (27/2) \ \lambda_t^2\alpha_s
(\zeta(3)/\pi^2)^2  T}
and
so
$\bar D \Gamma_y/v_w^2\gtap 2/v_w^2$ and the assumption
that this rate is fast is self consistent.
The next largest Yukawa
coupling is the bottom quark's. Including scattering from this Yukawa
coupling
would give corrections to our results of order \eqn\gammb{\sim\bar D
\left((27/2)
\lambda_b^2\alpha_s (\zeta(3)/\pi^2)^2  T\right)/v_w^2.} We
assume that the ratio $\tan\beta$
of Higgs
expectation values is not  unnaturally large, and so scattering due to
the bottom and other Yukawa couplings
may consistently
be neglected for
$v_w\gtap 10^{-2}$.

For the anomalous fermion number violating rates we take
\eqn\sphalrates
{\Gamma_{ws}=6\kappa\alpha_w^4 T\ ,\qquad
\Gamma_{ss}=6\kappa'{8\over3}\alpha_s^4 T\ ,} where $\kappa,\kappa'$
are
unknown parameters
usually assumed to be of order one. Thus  the weak sphaleron rate may
safely be
taken to be
slow provided \eqn\slowws{\kappa/v_w^2\ltap 10^{4}}  and  the strong
sphaleron
rate is fast if
\eqn\fastss{\kappa'/v_w^2\gtap 5\ .} In our computation of the baryon
asymmetry
we will
approximate
the strong
sphaleron rate as fast and the weak sphaleron rate as slow.

What we set out to compute was not the Higgs density in the symmetric
phase but
the
total baryon number density left inside the bubble. We now turn the
weak
sphaleron
rate   on, assuming it has a negligible effect on particle densities
(eq.~\slowws\
is valid), however it provides   the only source for   net baryon
number.
We  thus take $\rho_B$, the baryon number density,  to be a function
of $\bar
z$
satisfying
\eqn\barrate{0=D_q\rho_B''-v_w\rho_B'-\Theta(-\bar z
)n_F\Gamma_{ws}n_L(\bar z)\ ,} where
  $n_L$ is the total number density of left handed weak
doublet fermions, $n_F=3$ is the number of families, and we have
assumed that
anomalous baryon
number creation takes place only for $\bar z<0$ (the symmetric phase).
Eq.~\barrate\ has
solution
\eqn\barratesol{\rho_B(\bar z)=-{3\Gamma_{ws}\over
v_w}\int_{-\infty}^0
n_L(\bar
z) d\bar z-
{3\Gamma_{ws}\over v_w}
\int_0^{\bar z}dz'\Theta(-z')n_L(z')\left(1-e^{v_w(\bar
z-z')/D_q}\right)\ ,}
which is a constant
for $\bar z>0$ and vanishes as $\bar z\rightarrow -\infty$. Thus,  up
to
 corrections   of order
$\Gamma_{ws} \bar D/v_w^2$, the baryon density inside the bubbles
of broken phase is simply proportional to the integral of   $n_L$ in
the symmetric phase.

We now return to eq.~\masteq, in order to compute $n_L$. As pointed
out by
Giudice and
Shaposhnikov \giudice, if we   use eq.~\counting\ we will find the
answer
is zero in the
limit
$\Gamma_{ss}\rightarrow\infty$, so we need to compute the
$\CO(1/\Gamma_{ss})$
corrections to
particle densities\foot{
In this limit, one should also
include the contribution of sources for
conserved charges: $B-L$, $\ldots$. Local densities of conserved
charges  are also
generated by the scattering of
particles on the moving wall through charge separation.
As discussed in section \S 2.1,
these sources are subleading, however   they do not  suffer from the strong
sphaleron suppression.}.
   We will assume
$\Gamma_y\gg \Gamma_{ss}$, ($\kappa'\ltap 7$) and take
\eqn\qtsolii{\eqalign{Q= &H\left({ k_Q(9k_T-k_B)\over k_H(k_B+9
k_Q+9k_T)}
\right)+\delta_Q+\CO(1/\Gamma_y)\cr
T=& -H \left({k_T(2k_B+9k_Q)\over k_H(k_B+9 k_Q+9k_T)}
\right)+\left({k_T\over k_Q}\right)\delta_Q+\CO(1/\Gamma_y)\ .}}
Substituting
 these values into eq.~\masteq, we find
\eqn\qtii{\eqalign{
0=&D_q(Q''+T'')+v(Q'+T')-
   3\Gamma_{ss}\left({2Q\over k_Q}-{T\over k_T}+{9(Q+T)\over
k_B}\right)\cr
=&\left({-k_B(k_Q+2k_T)\over
k_H(9k_Q+9k_T+k_B)}\right)\left(D_qH''-v_wH'\right)
   -3\Gamma_{ss}\left({k_B+9k_Q+9k_T\over k_B k_Q}\right)\delta_Q\cr
 &+\CO(1/\Gamma_{ss},1/\Gamma_y)\cr
\Rightarrow\delta_Q=&\left({D_qH''-v_wH'\over\Gamma_{ss} }\right)
\left({-k_B^2k_Q(k_Q + 2k_T)\over3k_H( 9k_Q + 9k_T+k_B )^2}\right)+
\CO(1/\Gamma^2_{ss},1/\Gamma_y)\ . }}
We   now solve algebraically for $n_L=Q+Q_{1L}+Q_{2L}$ using
eqs.~\algebra,
\qtsolii, and \qtii, and find \eqn\nlsol{\eqalign{n_L=&5Q+4T\cr
=&\left({5k_Q+4k_T\over
k_Q}\right)\delta_Q+\left({9k_Qk_T-8k_Bk_T-5k_Bk_Q\over
k_H(k_B+9k_Q+9k_T)}\right) H\ .\cr}} If we use eq.~\counting\ we find
\eqn\leftdensi{n_L=
7 \delta_Q={-1/56}\left({D_qH''-v_wH'\over\Gamma_{ss} }\right)\ ,} so
the
baryon density is
proportional to $\Gamma_{ws}/\Gamma_{ss}$, and is only sensitive to
the ratio
of $\kappa/\kappa'$, provided  eqs.~\slowws\ and \fastss\ are
satisfied. This reduces the uncertainty in the baryon asymmetry since
estimates for both $\kappa$ and $\kappa'$ vary by several orders of
magnitude but we expect the ratio to be approximately one.
The result that
$n_L$ and the baryon density are suppressed by a factor of
$1/\Gamma_{ss}$
does not hold if
one considers modifications to eq.~\counting\ due to higher order
corrections
\giudice\ or due
to the contributions of nondegenerate squarks.
The cancellation
which makes the first term of eq. \nlsol\ dominate the second term
no longer occurs when nondegenerate masses are considered.
However  examination of eqs.~\qtii\ and \nlsol\ shows that
corrections from this lack of cancellation are
negligible unless either
$\kappa'/v_w^2\gtap 10^3$, or    some
squarks are
 not much heavier  than $T$. Note that there can be significant
enhancement of
the baryon density
if, for example, only the top squark is  light in the symmetric
phase, as we
will
discuss at the end of this section.
With all squarks heavy, our final answer for the baryon to entropy ratio
in the
broken
phase,
 combining eqs.~\sol,
\approxampl,
\barratesol,
\qtii, and
\nlsol\ is
\eqn\finsol{\eqalign{
{\rho_B\over s}=&\,-\,
\left({3 {\cal A}\Gamma_{ws}\over56s\Gamma_{ss}
}\right)\left(1-{D_q\over \bar
D}\right)\cr
\approx&\, - \, 3.5\times 10^{-5}\, \gamma_w v_w \
\left({\kappa\over\kappa'}\right)
 \  \CS
\cr}
} with
\eqn\sfactor{
 \CS \ = \
{(\gamma_{\tilde s} +\gamma_{\tilde h}) w \over
(\Gamma_m + \Gamma_h) \gamma_w v_w T^2 }\ \left({1-e^{-2
w\sqrt{\tilde\Gamma\over\bar D}}
\over w T}\right) \ .
}

We have taken the entropy $s$ to be
$s=(2\pi^2g_*/45)T^3=55.1\  T^3\quad (g_*=125\ 3/4)\ $ and we have
made explicit the dependence on the velocity $v_w$ and the thickness
of the
wall $w$.

The factor $\CS$ is a dimensionless number, function of the
supersymmetric
parameters $\mu$, $\sin \phi_B$, ${\tilde m}_{1,2}$, $A$, $\sin
\phi_A$,
 $\tan \beta$ and $\Delta \beta$, the total  variation of $\beta$ in
the wall,
as well as a function of the known gauge and top Yukawa couplings
and $W$ and $Z$ masses. In short, $\CS$ is a  concise representation
of
 the dependence of the baryon asymmetry produced on the yet unknown
supersymmetric
 parameters of the SSM.

To compute the baryon asymmetry, we need to compute the factor in
parenthesis
in Eq. \sfactor. This factor has its origin in the mechanism which
transports
the \cp\ violating asymmetries in front of the wall.
If this transport is efficient, the
answer should become independent of the wall thickness $w$. Indeed,
using our estimates, $100/T$,  for $\tilde D$ given in \diffbar\ and
our estimates for $\tilde\Gamma$ in \effdconst\ and \hyper,
we find $2 \sqrt{\tilde\Gamma/\tilde D} \simeq 5 \times 10^{-3}
\sqrt{7 \sin^2\beta + 1} \ T$, a value fairly
insensitive on the supersymmetric parameters.
Hence, unless the wall thickness $w$ is anomalously
large,\foot{Typical estimates
for
$w T$ range between $10$ and $100$.}
we find the factor in parenthesis to be equal to $ 5 \times 10^{-3}
\sqrt{7 \sin^2\beta + 1} $ and,
at leading order, independent of the wall thickness.

The $\CS$-factor becomes, using
general expressions for the source terms  given in \Higgsinos,
\formHiggs\ and
 \squarks, \formsquarks, and our expressions for the relaxation rates
$\Gamma_m$ and
$\Gamma_h$ in \hyper,

\eqn\sgeneral{\eqalign{
\CS \, &\approx \, \
5.5 \times 10^{-3} \   { \mu \over T} { {\tilde m}_2 \over T} \sin
\phi_B
\Delta\beta  /
\sqrt{7  \sin^2 \beta + 1 }  \ \times  \biggl\{ \cr
 &\left( \ {\cal I}_{\tilde h}^\mu \,
(\, 1 \, +\,{{\tilde m}_1 \over {\tilde m}_2} {1\over 10}\, ) \, + \,
{\cal I}_{\tilde h}^{{\tilde m}_2}  \,+\, {\cal I}_{\tilde
h}^{{\tilde m}_1}
\, {{\tilde m}_1 \over {\tilde m}_2} \ {1 \over 10}\, \right) \ \sin
\phi_{B} \ + \cr
 &{}\qquad \qquad\left( \,\bigl[{\cal I}_{\tilde s}^{{\tilde m}_L} \, + \,
{\cal I}_{\tilde s}^{{\tilde m}_R} \bigr] \
\ 10 \ {A\over {\tilde m}_2} \right) \
\sin(\phi_B -\phi_A)
  \  \biggr\} \ \cr}
}

The first term in parenthesis represents the contribution of the
charginos and neutralinos while the second term represents the
contribution of the top squarks.

To try this formula, we use light neutralinos and charginos:
$\mu = {\tilde m}_2 = 2{\tilde m}_1 = 50$ GeV,
relatively
heavy
squarks; ${\tilde m}_{L,R} = 150$ GeV, and take $A=50$ GeV. We need to know
the ratio  of these
masses over the transition temperature $T$. The latter has its value
completely determined by the parameters of the theory; in our
analysis,
however,
it is a free parameter. As an indicative value, we choose $60$
GeV \foot{
Generically, one expects the temperature to be below the one in the SM
( $\sim 80-100 $ GeV),
as, in the MSM, the superpartners contribute to the effective
potential
in a manner which decreases the critical temperature.}.
For these values, we find ${\cal I}_{\tilde h}^\mu =
{\cal I}_{\tilde h}^{{\tilde m}_2}\approx 5.5\times 10^{-2}  $,
${\cal I}_{\tilde h}^{{\tilde m}_1}\approx 3.5\times 10^{-2} $  and
${\cal I}_{\tilde s}^{{\tilde m}_{L,R}} \approx 2.4\times 10^{-4}$.
the $\CS$-factor becomes
\eqn\snumeric{\eqalign{
\CS \,\approx&\, \left({5 \over 6}\right)^2 \ { \Delta\beta \over
\sqrt{1+7\sin^2\beta}}
\ \left( \, 0.032 \ \sin \phi_B \ + \
5 \times 10^{-3} \   \
\sin(\phi_B -\phi_A)    \, \right) \ 5.5 \times 10^{-3} \cr
\approx& \, \left( \, 1.8 \times 10^{-4} \ \sin \phi_B \ + \
 6.6 \times 10^{-6}\   \
\sin(\phi_B -\phi_A)    \, \right) \ \Delta \beta\ . \cr}
}
It is clear from the above equation that the squark
contribution is only significant in the limit $\phi_A \gg \phi_B$
or in the limit the charginos and neutralinos are heavy.

Gathering all the above information, we find the following results.
The largest contribution arises from light charginos and/or
neutralinos, in which case, the asymmetry can be as large as
\eqn\asymn{
{\rho_B \over s} \,\approx \,
- \ \gamma_w v_w \ \left( {\kappa \over \kappa'}\right) \
\sin\phi_{B} \  \Delta\beta \ 6.5 \times 10^{-9}.
}
The measured baryon asymmetry is $(4-7)\times 10^{-11}$.
So, electroweak baryogenesis is significant  provided that
\eqn\boundsusyone{
\left( {\kappa \over \kappa'}\right) \
\left|\gamma_w v_w \ \sin \phi_B  \ \Delta\beta \right| \geq   7.5
\times 10^{-3}
}
and is negative in sign. Let us discuss the magnitude of each term
separately.
$\kappa$ and $\kappa'$ are two not well-known parameters
characterizing the strength of the electroweak and strong anomalous
processes, respectively, however, their ratio is expected to be of
order one.  In the minimal standard model,
the wall velocity, $\gamma_w v_w$, is no smaller than
$0.02$\foot{This lower bound corresponds to the situation of maximal
damping of the motion of the wall in the plasma, that is, it
corresponds to the thin wall situation where mean free paths
are larger than the thickness of the wall, $w$. This lower bound is a
decreasing function of the Higgs and top quark masses; the specific
value $0.02$ has been computed following Ref. \dhlll\ for the values
$m_H \sim 65$ GeV and $m_t \sim 175$ GeV.} and is more likely of order
$0.1$\foot{This larger value accounts for thermal scattering within
the wall as, in the MSM, the wall thickness $w$ is typically larger
than the mean free paths of the $W$'s and  $Z$'s, $\tau_w$, and of
the top quark, $\tau_s$. Large uncertainties arise from our
imprecise knowledge of the ratios $\tau_{w,s}/w$.}
or larger\refs{\dhlll,\mlt}; although no
calculation has been done for SSM, it is  reasonable to assume
similar values. Finally, $\Delta\beta$ is the overall variation of the
ratio of the two Higgs expectation values $v_2$ and $v_1$. As we
argue in \S 5, its presence is an artefact of working at fourth order
in the mass, it can be removed at the cost of introducing additional mass
suppressions. From these
considerations, we infer an optimal bound on the \cp\ violating phase
$\phi_{B}$
\eqn\boundsusyh{ |\sin\phi_{B}| \, \geq \ 0.025 \ .
}
Only with this bound satisfied, is  electroweak baryogenesis
achievable in SSM with light
charginos and/or neutralinos and heavy and degenerate squarks.

In the case of
neutralinos and charginos which are heavier than
$T$  or $\phi_{A} \gg
\phi_{B} \sim 0$, only top squarks contribute to the asymmetry.  If all
squarks are degenerate, they must all be heavier than $\sim 150$ GeV in  which
case, the requirement becomes
\eqn\boundsusys{ |\Delta\beta \ \sin(\phi_{A}-\phi_{B})| \, \geq \ 0.65 \ .
} which leads to experimentally ruled out electric dipole moments.
Also, unlike the charginos/neutralinos case, $\Delta\beta$ must be non-zero
and its presence is not an artefact of our approximations;
$\Delta\beta$ can be significantly smaller than one. Clearly,
electroweak baryogenesis in SSM with
all neutralinos, charginos and squarks heavy is likely to be
incompatible with constraints from electric dipole moments.
We will discuss bounds \boundsusyh\ and \boundsusys\
along with their uncertainties in the last section.

These conclusions are altered considerably if, say,  the  left handed
bottom
squark and
left and right handed top squark masses squared are rather light, but
the other
squark
masses are heavy. (This mass pattern could be a result of
renormalization due
to
the large top Yukawa coupling near the Planck scale.) Then the
factor multiplying
$H$ in eq.~\nlsol\ does not vanish. If we take
\eqn\countingii{k_Q\approx18,\quad k_T\approx
9,\quad
k_B\approx 3,\quad k_H\approx 12\ ,} we have
\eqn\nosolii{n_L={27\over82} H\ ,}
\eqn\bardii{\bar D\approx 72/ T \ , }and
\eqn\finsolii{
{\rho_B\over s}=\, - \,\left({81 {\cal A} \bar D \Gamma_{ws}\over 82
v_w^2
s}\right) \
,} \ie,
$\rho_B$ is enhanced by a factor of $\sim 18 \bar D
\Gamma_{ss}/v_w^2$ over the
case
with no light squarks and fast strong sphalerons, and  is  sensitive
to the
weak sphaleron rate rather than the ratio of weak and strong
sphaleron rates.
After a few substitutions, we obtain
\eqn\finsolin{
{\rho_B\over s}\approx  - 1.5 \times 10^{-4} \ \CS / v_w
,}
where we have made use of the $\CS$-factor defined in \sfactor\ and
\sgeneral.
To obtain a numerical estimate, let us assume the values
$A \simeq \mu \simeq {\tilde m}_2 \simeq 2{\tilde m}_1 \simeq 50$ GeV, and
 ${\tilde m}_{L,R} \simeq T \simeq 60$ GeV. The  factors in
\sgeneral, are
 now
 ${\cal I}_{\tilde h}^\mu =
{\cal I}_{\tilde h}^{{\tilde m}_2}\approx 5.5\times 10^{-2}  $,
${\cal I}_{\tilde h}^{{\tilde m}_1}\approx 3.5\times 10^{-2}$  and
${\cal I}_{\tilde s}^{{\tilde m}_{L,R}} \approx 0.018$.
the $\CS$-factor becomes
\eqn\snumerii{\eqalign{
\CS \,\approx&\, ({5 \over 6})^2 \ { \Delta\beta \over
\sqrt{1+7\sin^2\beta}}
\ \left( \, 0.032 \ \sin \phi_B \ + \
0.093  \
\sin(\phi_B -\phi_A)    \, \right) \ 5.5 \times 10^{-3} \cr
\approx& \, \left( \, 1.8 \times 10^{-4} \ \sin \phi_B \ + \
 5 \times 10^{-4}  \
\sin(\phi_B -\phi_A)    \, \right) \ \Delta \beta \cr}
.}
This time the contribution of the top squarks --
the second term in  parenthesis, is potentially as significant as the
one of the light charginos or neutralinos.
Combining this result with \finsolin, we find

\eqn\rholight{
{\rho_B\over s}\approx  -  {\kappa \over v_w} \
 \left(  \Delta \beta \ \sin\phi_{B} \ 2.5 \times 10^{-8}
\ + \ \Delta \beta \ \sin(\phi_{B}-\phi_{A}) \ 7.5 \times 10^{-8}
\right) \ .
}
This is a significant contribution to the baryon-to-entropy ratio
provided that
\eqn\boundlight{
\left|  \Delta \beta \ \sin\phi_{B}\ + \
       3 \, \Delta \beta \ \sin(\phi_{B}-\phi_{A})
\right| \ \geq \ {v_{W} \over \kappa}   \  1.9 \times 10^{-3} \ .
}
and is negative in sign.

If we use the range
$0.1-0.3$ for the wall velocity $v_w$ and the range $0.1-1$ for
$\kappa$, and take masses for the superpartners which are optimal for
baryogenesis, we obtain the following   constraint on the magnitude
of the  phases $\phi_{A}$ and $\phi_{B}$
\eqn\boundlight{
\matrix{
{\rm light \ \ charginos/neutralinos/top \ squarks} &
\left|  \Delta \beta \ \sin\phi_{B}\right|   &
\geq \ 2\times 10^{-4} \ - 6
\times 10^{-3} \cr {\rm light \ top \ squarks/charged \ higgs} &
\left| \, \Delta \beta \ \sin(\phi_{B}-\phi_{A})\right|
& \geq \ 7 \times 10^{-5} \ - 2 \times 10^{-3} \cr} }
We emphasize  that in the term contributed by the charginos and
neutralinos,
$\Delta\beta$ will  not   be present at higher order in the mass expansion.
By taking the top squarks to be light we obtain a possible two-order of
 magnitude
enhancement in $\rho_B/s$ over the situation with all squarks
degenerate and heavier that $T$
(cf. \boundsusyone-\boundsusys).

We also obtain qualitatively similar results to eq.~\finsolii\ if
strong
sphalerons are slow, \ie\ eq.~\fastss\ does not hold. Our formulae
are also
radically modified if weak sphalerons are sufficiently fast and/or if
the wall
velocities are so slow that eq.~\slowws\ is violated. Then
most of our simplifications of the rate equations, such as the
neglect of
leptons,
are invalid.  We then expect the final answer for
$\rho_B$ to   be  insensitive to the sphaleron rates, being determined
by near-equilibrium physics.

\newsec{Baryon density in the Two Higgs model}

We can now easily solve for the baryon density in the two Higgs
model since the particle transport equations  are very similar to
those in the SSM. Eq.~\masteq\ is unchanged, if we take the squark
and Higgsino contributions to be zero, $\gamma_{\tilde s}$ to be the
source for axial top number due to the top quark, $\gamma_q$
 and we substitute $\gamma_H$,   the source for Higgs number due
to the Higgs particles, for
$\gamma_{\tilde h}$. Finally, in the case of two(one) light Higgs,
the statistical factors become
\eqn\countinghh{k_Q\approx6,\quad k_T\approx
k_B\approx 3,\quad k_H\approx 8(4)\ ,}
and the effective diffusion constant, from \effdconst\
\eqn\bardhh{\bar D\approx {96 \over T}\ \left( {88 \over T}\right) \
. }

In Ref.~\us\ we computed $\gamma_q$ to be
\eqn\fromus{
\gamma_q ({\vec x},t) \, \simeq \,  -{ N_c\over 2 \pi^2}\,
\gamma_w v_w T\ |m_t|^2
\partial_z \theta \ + \ {\cal O}\left(v_w^2, (\tau/w)^2\right),
}
where $m_t(z)$, $ = |m_t(z)| \, e^{i \theta(z)}$,
is the space-dependent mass of
the top
quark expressed
in the wall frame.
To find $\gamma_H$, we need to track the evolution of the Higgs
number
carried by the Higgses $H_1$ and $H_2$ as they evolve in the
background of the
wall.
The space-dependent mass matrix is, in the basis $H_1$, $H_1^*$,
$H_2$ and
$H_2^*$,
\eqn\honehtwo{
\mtwoh \,=\,
\pmatrix{ \ldots&  m_{11}^2 e^{-i \theta_{11}} & \ldots & m_{21}^2
e^{-i
\theta_{21}}\cr
 m_{11}^2 e^{i \theta_{11}} &\ldots &  m_{12}^2 e^{i \theta_{12}} &
\ldots \cr
 \ldots & m_{12}^2 e^{-i \theta_{12}} &\ldots &  m_{22}^2 e^{-i
\theta_{22}}  \cr
 m_{21}^2 e^{i \theta_{21}} &\ldots &  m_{22}^2 e^{i \theta_{22}} &
 \ldots \cr}
.}
We only displayed entries which violate Higgs number as they are the
ones which
control
the charge generation as Higgs particles flow across the wall.
The Higgs number charge operator takes the form
\eqn\chargestop{
{\hat Q}_{H} \,=\, {\rm Diag}\bigl(\,1,-1,1,-1\, \bigr)
.}
The analysis follows the steps of the one of the stop axial charge
generation;
in
particular,
Eq. \jptwo\ and \jmtwo\ are directly transposable.
We obtain
\eqn\Higgsource{
\gamma_H ({\vec x},t)\, = \, \gamma_w v_w \ {1 \over 2 \pi^2} \
\times \ \bigl(
\J_{H}^{m_1}
\
 {\cal I}_{H}^{m_1}\, + \, \J_{H}^{m_2} \ {\cal I}_{H}^{m_2} \bigr)
\ + \ \  {\cal O}\left( v_w^2, (\tau/w)^2\right)
,}

with
\eqn\jhiggsone{\eqalign{
\J_{H}^{m_1}\ =& \ 2 \partial_z \theta_{11}\  m^4_{11}/T \ + \
\partial_z \theta_{21}\  m^4_{21}/T \  + \
\partial_z \theta_{12} \  m^4_{12}/T \cr
\J_{H}^{m_2}\ =& \ 2 \partial_z \theta_{22} \  m^4_{22}/T \ + \
\partial_z \theta_{21} \  m^4_{21}/T \  + \
\partial_z \theta_{12} \  m^4_{12}/T \, .\cr}
}
The   function  ${\cal I}_{H}^{m_i}$ is identical to the one computed
for the
squark,
${\cal I}_{\tilde s}^m$ in Eq. \formsquarks.
The damping rate is set by weak interactions, our estimate is $\tau_H
\sim
25/T$.
We choose for the on-shell masses $m_1$ and $m_2$ of the propagating
Higgses
the zero-momentum contributions
that the Higgs particles receive from plasma interactions in both
phases: $m_1
\sim m_2
 \sim T/3$ \susyphase.

With these values,  ${\cal I}_{H}^{m_i} \approx 0.25$ and,
within our approximations,

\eqn\sourceH{\eqalign{
\gamma_H ({\vec x},t)\, \simeq& \, \gamma_w v_w \ {1  \over 4 \pi^2} \
\times \cr
 &\left\{  \partial_z \theta_{11}\  m^4_{11}/T \ + \
    \partial_z \theta_{22}\  m^4_{22}/T \ + \
 \partial_z \theta_{21}\  m^4_{21}/T \  + \
 \partial_z \theta_{12} \  m^4_{12}/T \right\} . \cr}
}

 Combining eqs.~\sol,
\approxampl,
\barratesol,
\qtii, and
\nlsol, the baryon to entropy ratio is
\eqn\finsolhh{\eqalign{
{\rho_B\over s}=&\,-\,
\left({3 {\cal A}\Gamma_{ws}\over56s\Gamma_{ss}
}\right)\left(1-{D_q\over \bar
D}\right)\cr
\approx&\, - \, 2.4\times 10^{-5}\, \gamma_w v_w \
\left({\kappa\over\kappa'}\right)
 \  \H
\cr}
} with most of the parameter dependence contained in the $\H$ factor
\eqn\hfactor{
 \H \ = \
{(\gamma_q +\gamma_h) w \over
(\Gamma_m + \Gamma_h) \gamma_w v_w T^2 }\ \left({1-e^{-2
w\sqrt{\tilde\Gamma\over\bar D}}
\over w T}\right) \ .
}
 From \effdconst, we compute $\bar\Gamma \,=\,
0.11\,(\Gamma_m+\Gamma_h)$.
We estimate $\Gamma_m \approx \lambda_t^2 T/21$ and we parametrize
$\Gamma_h \approx \lambda^2 T/140$ with $\lambda$, an
undefined parameter function of the
Higgs quartic couplings.
As in the SSM,  charges diffuse a long distance in front of the wall
$\bar D \gg \bar \Gamma$ and the term in parenthesis in Eq. \hfactor,
is largely independent of the wall thickness $w$. We  find
\eqn\hhfactor{
 \H \ = \
{(\gamma_q +\gamma_h) w \over
\sqrt{(\Gamma_m + \Gamma_h)/T} \gamma_w v_w T^3 }\ 7 \times 10^{-3}\ .
}
Without going into a difficult study of the vast parameter space of
the
two Higgs models, we can obtain a fair estimate of the above quantity
by neglecting the Higgs contributions to the source and to the rate,
for the following reasons.
The source $\gamma_H$ written in \sourceH,
is a linear combination
of terms of the form $m^4_{ij}/T \partial_z \theta_{ij}$.
These terms all violate Higgs number,
hence, they are proportional to the quartic self-couplings
$\sim \lambda^2 (\Delta\theta_{ij}/w) T^4 $ and are smaller than
the contribution from the top quark $\sim \lambda_t^2 (\lambda\Delta
\theta_{ij}/w) T^4 $,
unless the Higgs sector is
strongly coupled.
Similarly, we expect $\Gamma_m > \Gamma_h$.
Under these assumptions,
\eqn\hhfactorn{
 \H \ \approx \ - 10^{-2} \ \Delta \theta
\
} and
\eqn\finsolhhn{
{\rho_B\over s}\, \approx\,  \gamma_w v_w \
\left({\kappa\over\kappa'}\right) \ \Delta\theta \ 2.5\times 10^{-7}
\ .}  Choosing the illustrative value $\gamma_w v_w \sim 0.3$, this
baryon per entropy ratio is significant provided that
\eqn\phasehh{
\left({\kappa\over\kappa'}\right) \ \Delta\theta \,\geq \, 7 \times
10^{-4}
\ .}

\newsec{Outlook}
\subsec{Accuracy of present computations of the baryon asymmetry}
We now look back on the many approximations and uncertainties present
in our analysis.

{\it -- Approximations -- }
\lfm{1.}
For the purpose of solving Majorana, Dirac and Klein-Gordon
equations, we
performed
an expansion in powers of $\mass(x,t)$. The benefit was to work
analytically
and to express the answer as a sum of \cp\ violating invariants.
The convergence of this expansion has been discussed and established
in the discussion following Eq. \hanalytic.
We have further approximated the density matrices describing particle
distributions with Bose-Einstein and Fermi-Dirac distributions for
on-mass shell particles in the unbroken phase.
In ignoring the non-equilibrium component of the
distribution, we are ignoring corrections of order $v_w^2$
(cf. discussion following Eq. \fermidirac).
In assuming on-mass shell particles  in the unbroken phase,  we are
ignoring corrections of order $(m/T)^2$ and $(gv/T)^2$. Some
particles, such as the squarks, are expected to  have $SU(2)\times
U(1)$ symmetric contributions to their masses which may be larger
than the
critical temperature  ($T_c\sim$ 50--100 GeV). However, as discussed
in \S 5.2,
heavy
particles do not  contribute significantly to baryogenesis.
So, at best, we
expect that accounting for the full mass dependence yields numerical
corrections
of order one. One exception is that for the neutralino and chargino
contribution, when we work to lowest nontrivial order in the masses
we obtain a
result proportional to $\Delta \beta$--the change during the
transition in the
angle specifying the ratio of the Higgs vevs. There is no reason to
expect this
suppression factor to persist at higher orders  in a mass expansion.

\lfm{2.}
We defined our sources
$\gamma_Q$  in a layer of a size $\tau$, the coherence time.
To postpone recourse to numerical methods, we assumed $\tau$ to be
smaller
than the wall thickness $w$ and ignored corrections of order
$(\tau/w)^2$.
This is a very good approximation for strongly interacting particles
but
not necessarily for weakly interacting particles for which $\tau$ is
in the
range $(20-30)/T$ while the wall thickness $w$ can span the interval
$(10-100)/T$. Only a precise calculation of these two quantities can
decide
 the quality of this approximation.
The largest  $(\tau/w)^2$ corrections are contained in the  factors
 ${\cal I}_{{\tilde h},{\tilde s}}$. As explained in \S~2, we expect
these  factors, which at most increase linearly with $\tau$, to
``saturate" for $\tau \gg w$, at about their  values at $\tau \simeq
w$.
For this reason, we do not expect higher order terms to bring
large corrections to our analysis.

\lfm{3.} We have made a number of simplifications of the   equations
describing particle transport and number changing processes. First, we
assumed that deviations from thermal equilibrium were sufficiently
small to
allow us to describe particle distributions in terms of local chemical
potentials and to make a diffusion approximation to transport
processes. We
expect this assumption to be quite good in the weakly interacting
models
considered.
We simplified our treatment of diffusion in neglecting the
finiteness of the speed of sound, that is, we worked at leading order
in an expansion in $v_w/c_s$ where, $c_s = 1/\sqrt{3}$.
Our choice
of the magnitude of the wall velocity is such that it
is a fair approximation.  Should the wall velocity approach or be
larger than the speed of
sound,  diffusion is not a good approximation to   transport and our
computations are invalid. An improved calculation which covers large
wall
velocities has yet to be developed. We made a severe approximation  by
simplifying the wall shape (eqs.~\approxsource\ and
\approxdecay), which we expect to give an $\CO(1)$ estimate of the
true
solution. We also made   assumptions about the approximate rates of
strong and
weak sphalerons and the wall velocity, (eqs.~\slowws\ and \fastss),
and
simplified our equations by assuming that the interactions
proportional to the
top
Yukawa coupling were in thermal equilibrium. The size of corrections
from
these
assumptions depends on how well the inequalities \slowws\ and \fastss\
are
satisfied.
We also assumed similar diffusion constants for all quarks, an error
of order
$\alpha_w^2/\alpha_s^2\sim 10\%$. In fact this approximation for the
quark
diffusion constants is of very small
numerical significance since   diffusion is
actually dominated by the weakly interacting Higgs, which provides a
local
source for axial top number far from the bubble wall.
We also gave approximate
estimates for the statistical factors
$k_i$ defined by eq.~\thermaldensities--here we expect corrections of
order a
few
percent for light particles. The  corrections to the quark
statistical factors
are important if
$\kappa'/v_w^2 \gg\CO(10^2)$ since   from  eqs.~\barratesol, \nlsol\
and
\leftdensi\ we see that they give the only  contribution to the baryon
asymmetry
which is not suppressed by the strong sphaleron rate.
\lfm{4.} In most cases, our CP violating particle sources are
dominated by the large CP
violation
in the transmission of  low momentum particles over a distance $\Delta$
whose wavelength is comparable to  $\Delta$.
 For these particles,  kinetic theory
starts to break down, giving corrections of
$\CO(1)$ to our treatment.
\lfm{5.} We have neglected the effects of long range gauge fields,
which in
general have an $\CO(1)$ effect on the baryon density
\refs{\khleb-\clinescreen}.
\lfm{6.} We have not included the contributions of the
transport of conserved charges (such as B-L) to the baryon
asymmetry. Such effects are higher order in wall velocity and masses,
but may not suffer from the strong sphaleron suppression. We expect
inclusion of such effects to
change our results by at most $\CO(1)$.
 
 {\it -- Uncertainties -- }

Uncertainties in our estimate of the baryon asymmetry reflect not only
the approximations above but also, and dominantly, our poor knowledge
of
certain parameters. Those are the coherence times
$\tau_{{\tilde s},{\tilde h}}$ and $\tau_{H,q}$, the diffusion
constants $D$,
  the
reaction rates $\Gamma_{y,h,m}$ and the parameters $\kappa$ and
$\kappa'$ measuring the strength of the anomalous
processes. Fortunately in the most interesting situations, the latter
occur in
ratio
which significantly decreases the   uncertainty in the baryon
asymmetry.
Also, for large $\kappa$ the baryon asymmetry becomes insensitive to
$\kappa$. Much work is needed to refine the determination of these
parameters.
Only $\tau$,
$\kappa$ and
$\kappa'$ require understanding new
physics; the determination of  the other parameters faces only
technical
challenges. The parameters $v_w$ and $w$ describing the phase
transition are
also
left free, both because they are parameter dependent and because of
the
lack of accurate computations for the models under
consideration.

Finally we come to the main uncertainty, which is our lack of
knowledge of the
correct
model of weak symmetry breaking and of the  many new parameters
introduced by
any
extension of the MSM. It is our hope that computation of the baryon
asymmetry
can
provide a useful constraint on the weak symmetry breaking sector and
on
\cp-violation.

Because of the above uncertainties and also   because
of the approximations that we described earlier,   we believe that
the computability of the baryon asymmetry produced is reliable to an
order
of magnitude. It is with this caveat that we now present our
conclusions.

\subsec{Can the baryon asymmetry be produced in the SSM?}
Previous work on baryogenesis in supersymmetric models neglected the
enhancing effects of transport, and concluded that sufficient
\cp-violation for baryogenesis in supersymmetric models could be
marginally
consistent with  electric dipole moment constraints if one made
optimistic
assumptions about baryon number violating rates
in the phase boundary \refs{\dhss,\cn}, and
if chargino and neutralino masses were not too heavy. Our work shows
that
with reasonable assumptions about the rates of anomalous  processes,
sufficient
baryon asymmetry can be produced with  small
\cp\ violating phases of order
$10^{-(2-4)}$,   provided that the
top squarks and either the neutralinos or the charginos are light compared
with the
transition temperature. If only the top squarks are light, it is also
required   that the ratio of Higgs vevs is not fixed during the
transition,
while when the {\it inos} are light, the lowest order contribution
in  $m/T$ is
suppressed unless the ratio of Higgs vevs changes during the
transition. The
latter requirement implies that the effective theory during the
transition has
more than one light Higgs, which in turn means that at zero
temperature the
pseudoscalar and charged Higgs masses are not extremely heavy
compared with the
lightest Higgs mass. A light charged Higgs makes a potentially ruled
out
contribution to
$b\rightarrow s \gamma$
\ref\buras{ A.J. Buras and S. Pokorski,  Nucl. Phys. B424 (1994) 374,
hep-ph/9311345  } unless partially cancelled by a contribution from a
loop
containing light charginos and stops. We conclude that as far as
sufficient
\cp\ asymmetry is concerned, the SSM with some light superpartners
($\ltap 100$ GeV) is a good candidate for baryogenesis. With light
superpartners
and with \cp-violating phases of order
$10^{-(2-4)}$,  neutron and atomic electric dipole moments will be
below the current experimental bounds
\refs{\lowerbounds,\susyatomic}.
Furthermore, a large
fraction  of the  relevant range of masses for the
superpartners coincides with
the range to be probed by LEP II.

In a calculation assuming only one light Higgs, the MSSM, with minimal
superpartner content, has been shown to produce a phase transition
sufficiently
strongly first order to preserve the baryon asymmetry only when the
lightest
Higgs boson mass is less than 70 GeV and when at least some of top
squarks are
lighter than 110 GeV \susyphase. We do not expect these bounds to be
weakened significantly when the full parameter space for the Higgs
masses is
considered. Thus  the baryon number washout constraint on the MSSM
seems more
powerful than the constraint of sufficient
\cp-violation. It is however subject to the uncertainties in the
perturbative
calculations of phase transition parameters.

\subsec{Conclusions about the baryon asymmetry in the two Higgs
model, and
comparison with other calculations}

To summarize \S~4,   sufficient baryon asymmetry may easily be
produced
in a general model with two Higgs doublets and soft \cp-violation in
the scalar
potential, with \cp-violating phase as small as $\sim 7\times 10^{-4}$
(eq.~\phasehh). This result allows for a much smaller phase than most
earlier calculations in the two Higgs model. Here we explain how our
calculation differs the earlier ones.

The baryogenesis
mechanism of axially
asymmetric top quark  reflection from the bubble walls \ckn\  also
allowed a
small phase of order $10^{-5}$ in the two Higgs model, but only for
the
fine-tuned case where the bubble walls were thin, of order the
inverse top
mass. Refs. \refs{\ckn,\clinetwohiggs} concluded that when the
walls are thick, a completely negligible
\cp-violating  asymmetry  is produced in the symmetric phase from
top quark
reflection. However in those papers   several significant
effects are neglected, such as thermal scattering within the phase
boundary
which is especially important for thick walls. Thermal scattering
processes
tend to interfere with baryogenesis by destroying the quantum
coherence
necessary for
\cp-violation~\refs{\hs,\ghop}, but also can in some cases {\it
enhance} the
baryon asymmetry produced.  The enhancement comes about because
\cp-violating charge expectation values   within  the bubble wall
 can be converted to
\cp-violating thermal
particle distributions inside the wall by incoherent  thermal
scattering
processes, and these \cp-violating thermal
particle distributions can then diffuse into the symmetric phase,  where they
bias the relatively rapid anomalous weak processes towards producing net
baryon
number. We therefore find that the huge suppression of the top quark
contribution to baryogenesis,  found
in refs.~\refs{\ckn,\clinetwohiggs} when the bubble walls are thick,
is absent
when  thermal scattering and transport processes are considered.
Instead, we
find that the baryon asymmetry is not very sensitive to the width of
the boundary.

  Let us now compare our method of computation to
two alternative methods which have appeared in the literature.
For the case of thick boundaries an alternative method of
calculation of the particle distributions in the wall, which should
be about as
accurate as the thick wall approximations we made, would be to use
the method of linear response
\nref\linres{E.M. Lifschitz and L.P. Pitaevkii, {\it Statistical
Physics, part
2} Pergamon Press, Oxford (1980); A.L. Fetter and J.D. Walecka, {\it
Quantum
Theory of Many Particle Systems}, McGraw Hill, New York
(1971)}\refs{\dhs,\linres}, i.e. to compute the charge current
density produced
from an initial
\cp\ symmetric thermal particle distribution when  space-time
dependent
\cp\ violating terms in the Hamiltonian are turned on for a time
equal to the
thermalization time $\tau$, and then dividing by $\tau$ to get the
rate for
production of a \cp-violating charge in the phase boundary.   Such a
calculation
can be done diagrammatically, \eg by computing  the diagrams
considered in
ref~\ref\cpm{D. Comelli, M. Pietroni and  A. Riotto, preprint
DESY-95-109,
hep-ph/9506278
}  (which however does not contain a linear response calculation, as
in that
work the effects of a finite
$\tau$ are neglected). If one considers times longer than $\tau$ in
the linear response, including the damping terms in the quark
propagators which are generated by gluon exchange is essential.
Another method of calculation has been developed in Ref.
\ref\classic{
M. Joyce, T.  Prokopec and N. Turok, Phys.Rev.Lett. 75 (1995) 1695,
hep-ph/9408339; preprint PUPT-1495, hep-ph/9410281;
preprint PUPT-1496, hep-ph/9410282.}.
It consists of writing a Boltzmann equation for a one-particle
distribution function which incorporates a \cp-violating force term arising
from the  \cp-violating space-dependent background. This
equation is then solved for the resulting departure from
thermal particle distributions, which is asymmetric between   particles and
their CP conjugates, and which are described in terms of chemical potentials.
The latter are, in turn, inserted in a rate equation to compute a baryon
asymmetry. It is not clear to us how to generalize this method to cover
the case where several species mix, as occurs with the neutralinos and
higgsinos in the SSM. In any case this method is semiclassical in nature and
only describes the regime of
$\tau$ much larger than particle wavelengths ($T\tau \gg 1$). In this
regime  our calculations also produce a semi-classical fall-off of the
\cp\ violating  sources (cf. Fig. 3 and Fig. 5), in qualitative agreement with
the  analysis of ref~\classic\foot{These authors coined the word ``classical
baryogenesis'' to describe their analysis. However  their ``classical force''
is not completely classical. That is, the spin dependent
\cp-violating term in their one particle Hamiltonian is $\propto \hbar
\partial_z \theta$, where
$\theta$ is the argument of the top quark mass.}. This calculation is
appropriate for weakly  interacting particles such as the $\tau$-lepton in the
two Higgs  model, but not for  strongly interacting particles whose mean free
paths are not much longer than their wavelengths and for which
decoherence effects are already perceptible.

It is now evident that transport processes, omitted in the original
thick wall calculations \refs{\dhss,\lsvt,\cknthick},
significantly enhance  the baryon
asymmetry produced during the weak transition.
In fact it has been suggested that the $\tau$-lepton plays a
leading role in baryogenesis due to its large diffusion constant
 \refs{\taurefs,\clinetwohiggs}. However axial top quark number is
also
efficiently transported, because the large top Yukawa coupling allows
axial
top number to convert to Higgs number, which is transported by weakly
interacting Higgs particles.
Another argument in favor of the tau lepton contribution to
baryogenesis
dominating that of the top quark is that the the axial top number
 tends to be
washed out by strong sphaleron processes.  In fact we find that this
suppression factor is only about $\sim 1/50$ for
$\kappa'$ of order one and wall
velocities  $v_w \sim 0.3$. Furthermore, even for arbitrarily fast
strong sphaleron rate, the strong sphaleron
suppression will never be more than about $10^{-3}$, due to the
nondegenerate
thermal masses of the quarks
\giudice.
Despite
the suppression factors for the top quark contribution,
we believe the tau is likely to be less important than the top for
baryogenesis
in two Higgs models, because the source for axial tau number    is
suppressed relative to the axial top source by a factor of
$\lambda_{\tau}^2/\lambda_t^2$, which is about
$10^{-4}$ unless $\tan\beta$ is large. In the SSM, it is only possible to avoid
 having sphalerons wash out the baryon number if $\tan\beta$ is relatively
small \susyphase, and so there is no significant  effect from the
tau or scalar tau.
\centerline{Acknowledgements}
This work was supported in part by the DOE under contract
\#DE-FG06-91-ER40614. The work of A. N. was supported in part by a
fellowship  from the Sloan Foundation. We gratefully acknowledge useful
conversations and correspondance with G. Bonini, M.B. Gavela, M. Joyce
and M.E. Shaposhnikov.

\nfig\figone{
(a) Amplitudes contributing to $J_+$.
(b) Amplitudes contributing to $J_-$.}

\nfig\figtwo{
(a) Contributions, to order $(\mass/T)^4$, to the transmission
amplitude $T$
of the neutralinos and the charginos.
(b) Corresponding contributions, to order $(\mass/T)^3$,
to the reflection amplitude $R$.
}

\nfig\figthree{
(a) The  factor ${\cal I}_{\tilde h}^m$ plotted versus $\tau T$.
${\cal I}_{\tilde h}^m$ contains kinematic
information on the propagation of the
neutralinos and charginos in the plasma.
(b) Its dependence on the mass eigenvalue $m$.
The dots are the result from numerical integration and the solid
lines are the fit \hanalytic.}

\nfig\figfour{
(a) Selected  contributions, to order $(\mass/T)^4$,
to the transmission amplitude $T$
of the squarks.
(b) Leading contributions, to order $(\mass/T)^2$,
to their reflection amplitude $R$.
}

\nfig\figfive{
(a) The  factor ${\cal I}_{\tilde s}^m$ plotted versus $\tau T$.
This  factor contains kinematic information on the propagation of the
squarks in the SSM and on the propagation of the Higgs particles in
two Higgs models.
(b) Its dependence on the mass eigenvalue $m$. The dots  are the
result from
numerical integration and the solid
line is the fit \squarksanaly.
}

\listrefs
\listfigs
%
%
	\def\epsfsize#1#2{\hsize}
	\vfill\bigskip\epsfbox{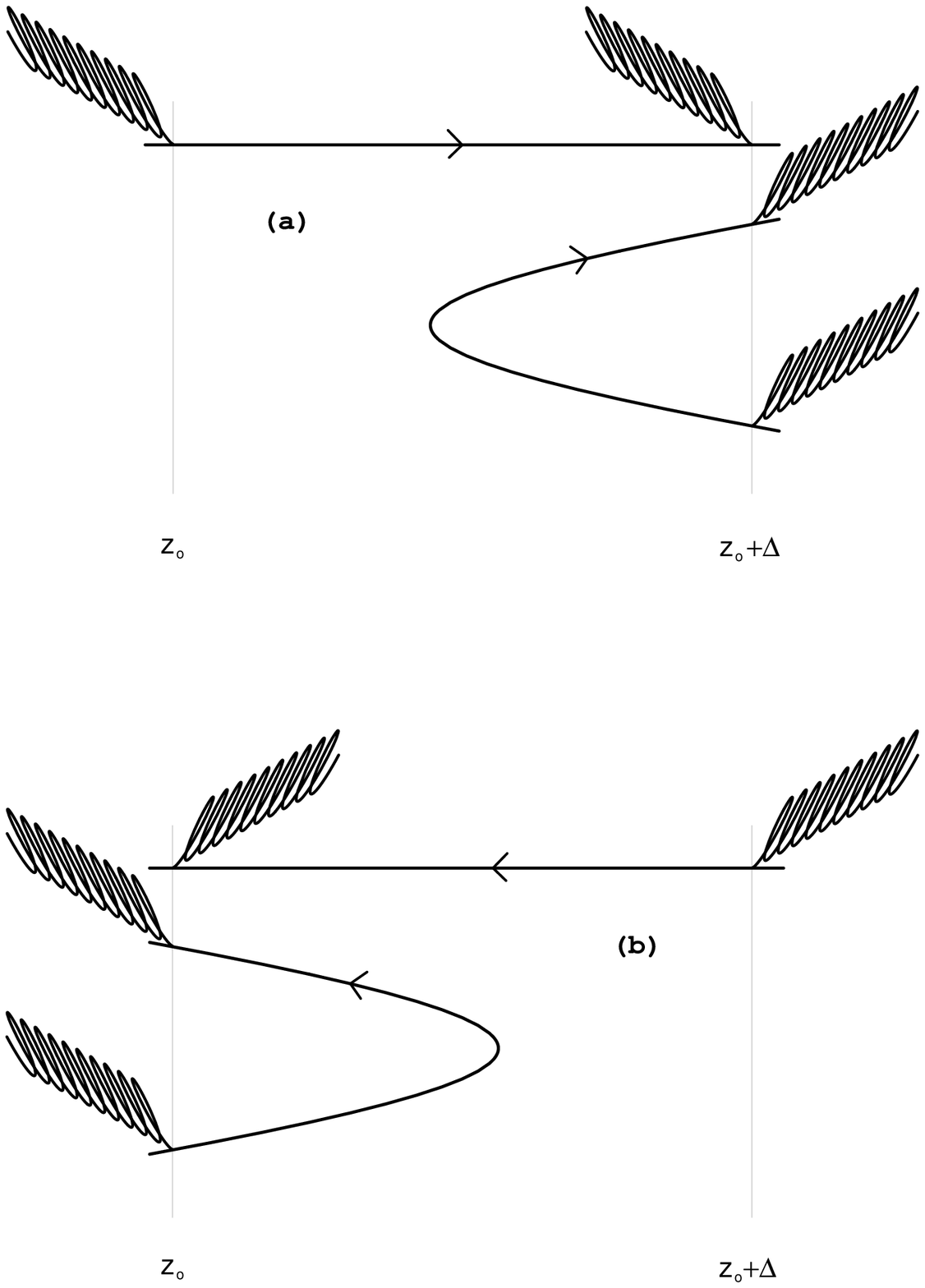}\vfil\eject	  %
	\vfill\bigskip\epsfbox{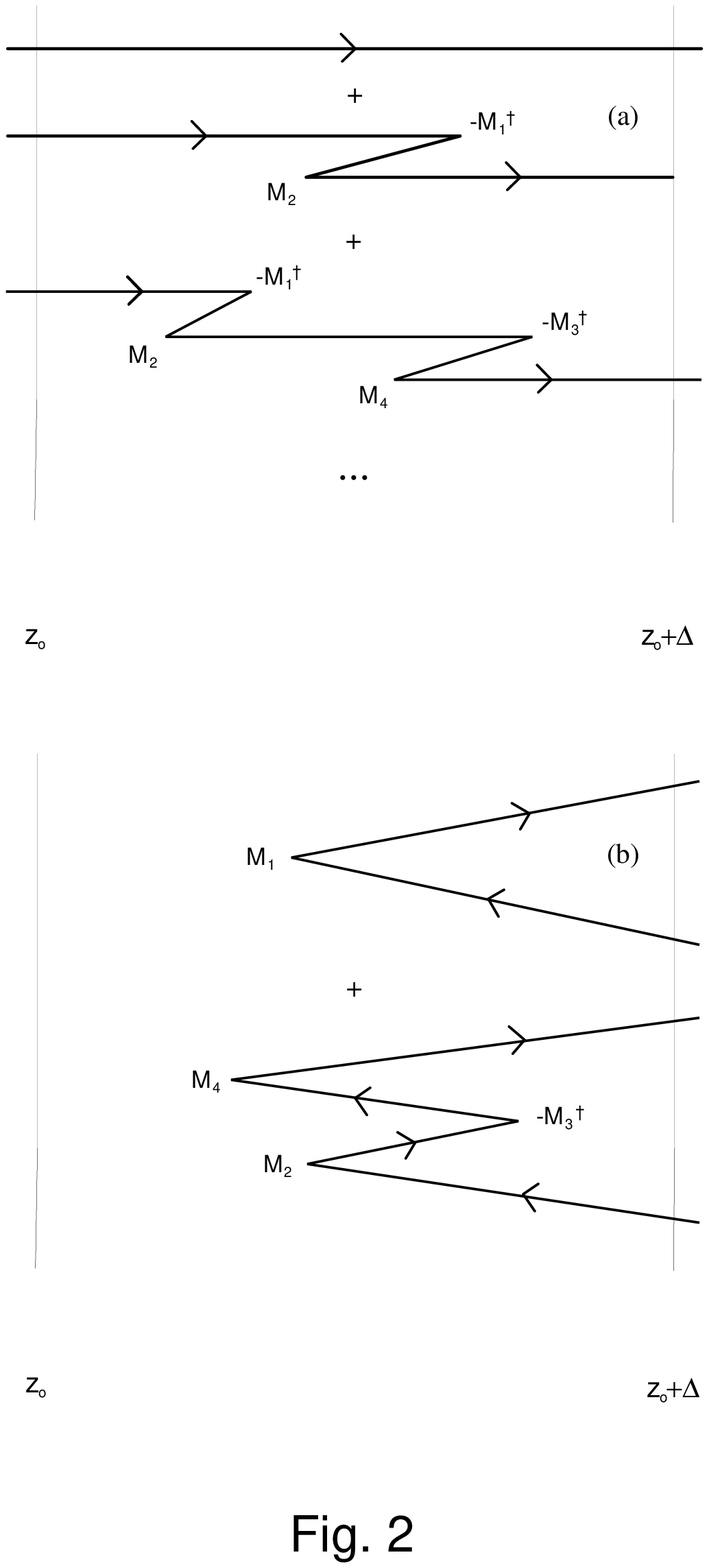}\vfil\eject	  %
	\vfill\bigskip\epsfbox{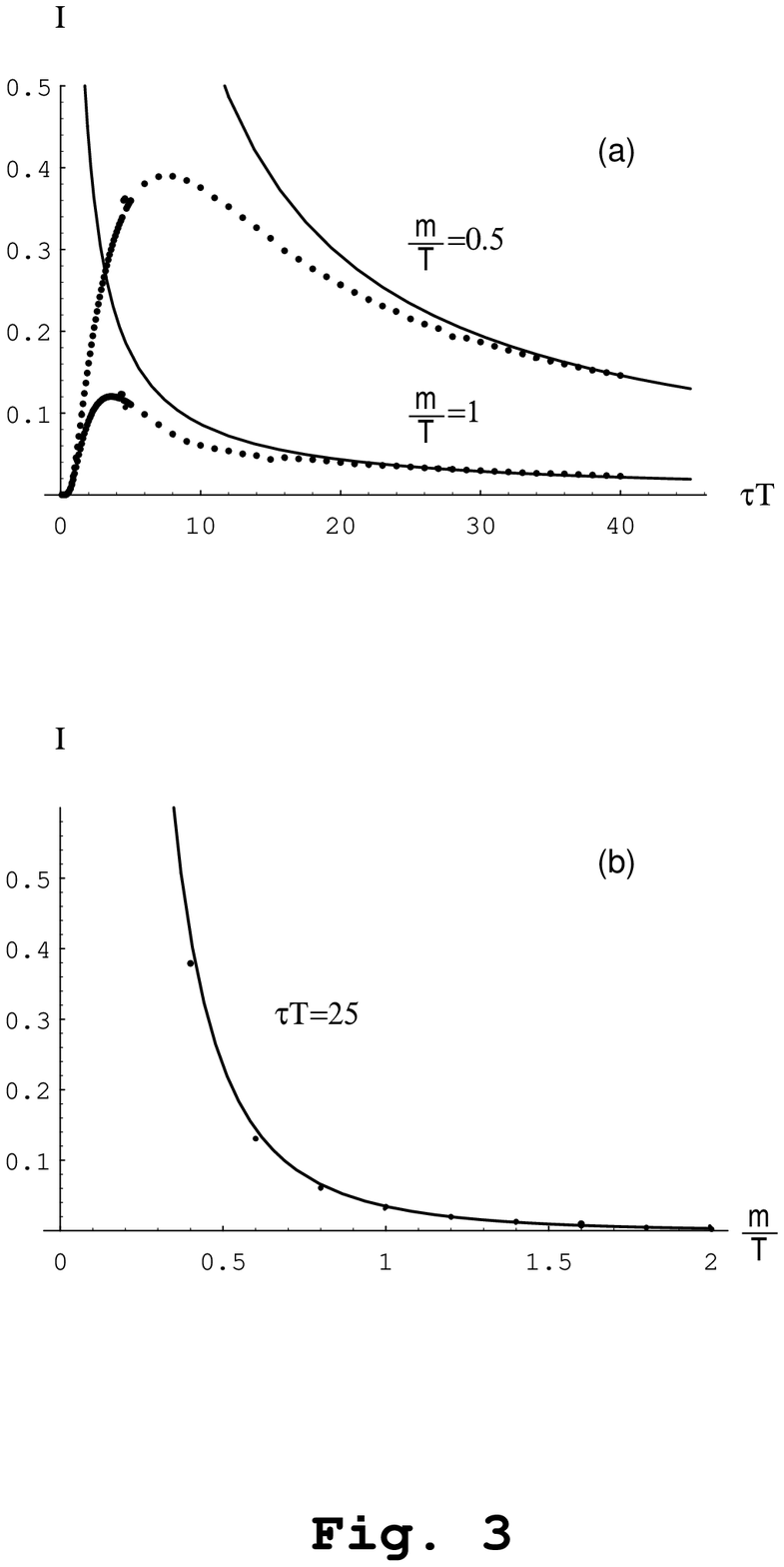}\vfil\eject	  %
	\vfill\bigskip\epsfbox{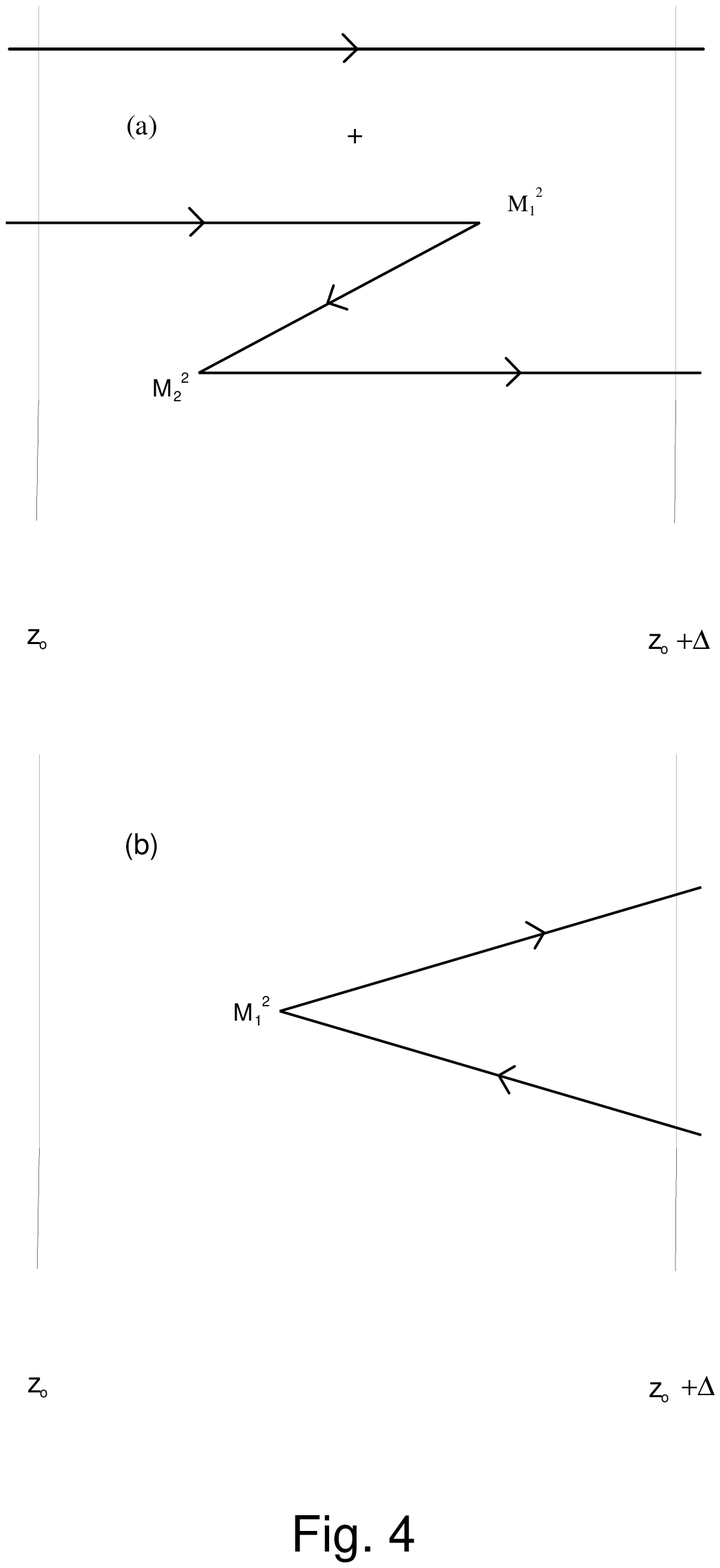}\vfil\eject	  %
	\vfill\bigskip\epsfbox{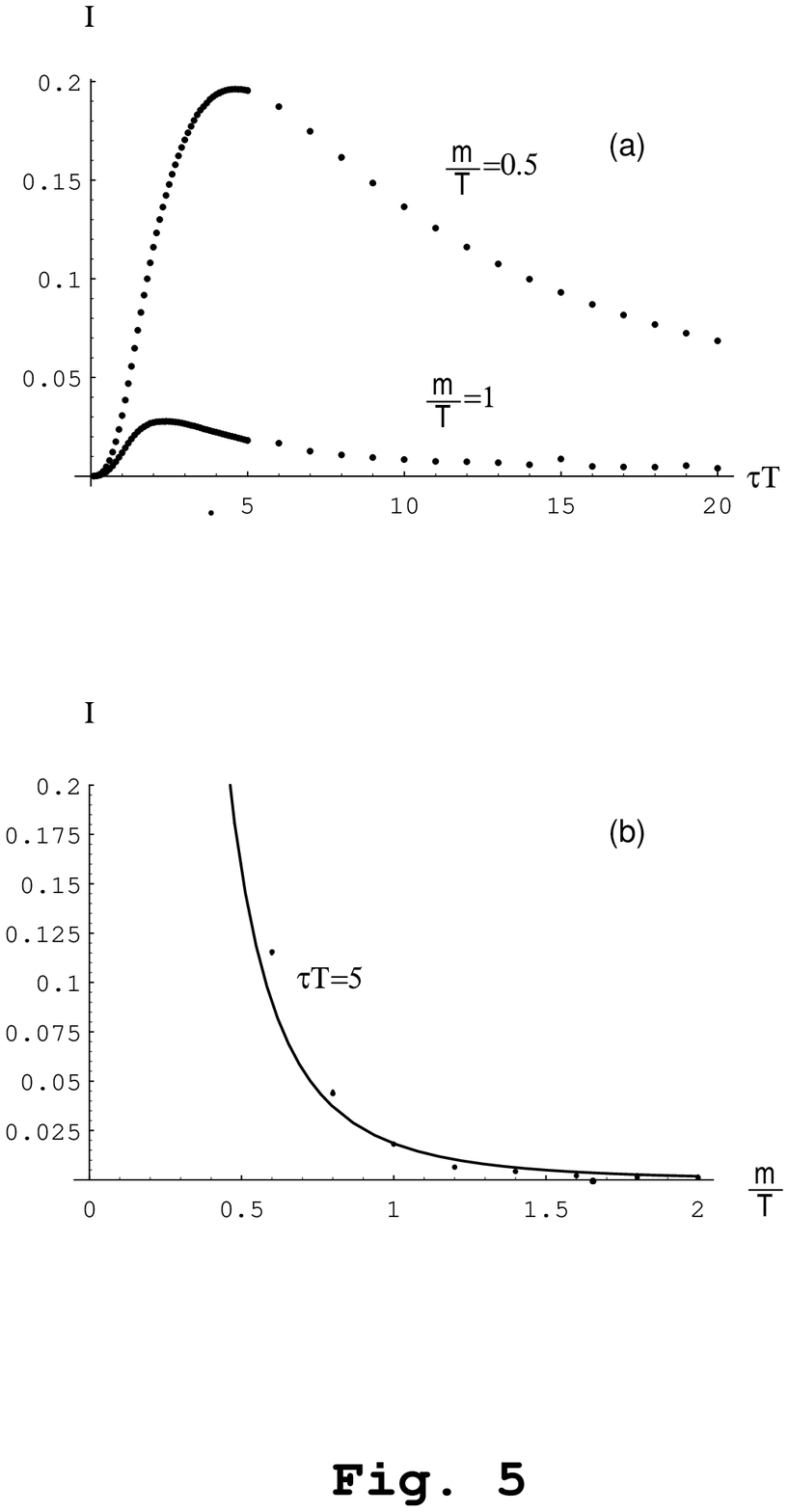}\vfil\eject	  %
%
%
%
%

\bye